# Machine Learning Applications to Computational Plasma Physics and Reduced-Order Plasma Modeling: A Perspective

Farbod Faraji*[1] and Maryam Reza*

*Plasma Propulsion Laboratory, Department of Aeronautics, Imperial College London, London, United Kingdom

**Abstract**: Machine learning (ML) provides a broad spectrum of tools and architectures that enable the transformation of data from simulations and experiments into useful and explainable science, thereby augmenting domain knowledge. Furthermore, ML-enhanced numerical modelling can revamp scientific computing for real-world complex engineering systems, creating unique opportunities to examine the operation of the technologies in detail and automate their optimization and control. In recent years, ML applications have seen significant growth across various scientific domains, particularly in fluid mechanics, where ML has shown great promise in enhancing computational modeling of fluid flows. In contrast, ML applications in numerical plasma physics research remain relatively limited in scope and extent. Despite this, the close relationship between fluid mechanics and plasma physics presents a valuable opportunity to create a roadmap for transferring ML advances in fluid flow modeling to computational plasma physics. This Perspective aims to outline such a roadmap. We begin by discussing some general fundamental aspects of ML, including the various categories of ML algorithms and the different types of problems that can be solved with the help of ML. With regard to each problem type, we then present specific examples from the use of ML in computational fluid dynamics, reviewing several insightful prior efforts. We also review recent ML applications in plasma physics for each problem type. The paper discusses promising future directions and development pathways for ML in plasma modelling within the different application areas. Additionally, we point out prominent challenges that must be addressed to realize ML's full potential in computational plasma physics, including the need for cost-effective high-fidelity simulation tools for extensive data generation.

## Section 1: Introduction

The importance of computational plasma physics to empower fundamental research and to aid engineering developments is well-established. The increase in computational power in addition to algorithmic advances have led to detailed plasma simulations over the past years that have revealed much about plasmas behavior, as well as their intricate underlying processes and interactions [1][2].

The next frontiers in plasma science and engineering, however, represent significantly challenging problems. On the fundamental science side, there are complex physics questions that have remained unresolved. These include the three-dimensional (3D) evolution of the plasma, interactions and coupling among various plasma instability modes, and the long-standing problem of particles' and energy transport across the magnetic field lines [3], including the spatiotemporal variations in instabilities' contribution to this phenomenon.

On the applied plasma physics side, several highly demanding, though crucially needed, engineering use cases have emerged, such as computer-aided design optimizations, state forecasting, and model-predictive control. In a broader sense, these applications form the underpinning capabilities for the transformation of the plasma industries toward digitalization and the integration of digital twins in the development processes of the plasma technologies [4].

Plasma modeling is faced with major challenges to address the above scientific and industrial needs. Most notably, on the one side, the highest fidelity conventional plasma models available today, i.e., kinetic particle-in-cell (PIC) codes [5][6], are prohibitively costly to investigate the multidimensional plasma phenomena across all the involved spatiotemporal scales in real-size plasma technologies for time durations of real-world significance [6]. This is true even in two dimensions, with the typical resources required for a 2D simulation being a cluster of thousands of CPUs and computation times of more than a year [7]. This enormous computational cost has precluded practical 3D PIC simulations, which will be costlier by at least an order of magnitude.

On the other side, less computationally demanding plasma codes, i.e., the "fluid" models, which solve a system of conservation equations for the plasma, are not self-consistent and predictive. This is because rigorous and generalizable closure models that can enable resolving the effects of missing kinetic and non-classical phenomena in these models without the need for a-priori assumptions or model tuning on experimental data are still to be developed [8].

---

[1] **Corresponding Author** (f.faraji20@imperial.ac.uk)

As demonstrated by the advances in the neighboring field to plasma physics, fluid mechanics – where issues rather similar in nature to those in computational plasma physics exist as well – machine learning (ML) and data-driven (DD) approaches can greatly help address the computational modeling challenges for both scientific and applied purposes.

It is important to note that data-driven modelling is not a new concept or one that is necessarily tied to machine learning. In fact, data-driven model discovery has long been a cornerstone of scientific advancement, with one prominent example being the Kepler's laws of planetary motion derived from Tycho Brahe's detailed astronomical observations. Today, the significant high-dimensionality and complexity of datasets make it impossible for the human brain to perform similar analyses unaided. This is one main challenge that has catalyzed the rise of machine learning to aid computational modelling and scientific discovery.

Machine learning is a subclass of Artificial Intelligence (AI) and comprises a broad array of algorithms, which aim to learn from data without the need for explicit mathematical models [9]. Machine learning is closely related to the data-driven framework, where the methods also seek to derive information/models from data. The term "data-driven" can be thought of as a general terminology encompassing both classic statistics/regression methods as well as modern generalized regression and ML approaches.

ML has a rich history that dates back to the 1950s and 1960s. In fact, two significant developments occurred in those years. One was the emergence of the concept of "cybernetics", as articulated by Wiener in 1965 [10], which aimed to mimic the thinking processes of the human brain. The other was efforts to develop "machines", such as the perceptron [11], with a focus on automating tasks such as classification and regression. The latter development has particularly laid the groundwork for modern machine learning, with ongoing relevance and impact in contemporary research. Research on ML and the broader field of AI saw a major drop in interest from the early 1970s to the late 1980s, following the criticizing report of Sir James Lighthill in 1974 [12] on the failure of AI to deliver on its grand promises. The interest in AI and ML, especially in neural networks, was reawakened in 1986 after the development of the "back-propagation" algorithm by Rumelhart et al. [13]. This was because the back-propagation algorithm enabled the training of neural networks with multiple layers. The links established by various authors [14][15] around the same time between machine learning and statistical mechanics also contributed to the resurged interest.

Machine learning continued to evolve and advance over the past decades [16], with significant interest and attention gathered in the past few years from the scientific and industrial communities toward incorporating the ML approaches in science discovery and technology development.

At the core of ML lies the concept of a "learning machine". A learning machine is an algorithm that iteratively learns patterns, structures, or relationships from data, allowing it to make predictions, classifications, or decisions based on new, unseen data. Cherkassky and Mulier [17] in 2007 defined the learning problem itself as "*the process of estimating associations between inputs, outputs, and parameters of a system using a limited number of observations* [16]".

ML algorithms can be categorized into three main types based on the nature of the learning process and the extent to which external supervisory information is available to the learning machine: **(1)** supervised learning (e.g., regression methods and neural networks), **(2)** unsupervised learning (such as, clustering techniques and dimensionality reduction methods), and **(3)** semi-supervised learning (most notably, generative adversarial networks and reinforcement learning). We will further discuss these various categories and some of the main algorithms belonging to each within subsection 1.1.

The application areas of ML for computational modelling of physical systems can be overall divided into two general problem types: **(1)** Capturing expensive physics cost-efficiently, which can consist of either accelerating high-fidelity simulation or developing low-cost reduced-order (surrogate) models for the phenomenon, process, or system of interest, **(2)** Discovery of the underlying dynamics, and discrepancy modelling, which aims to bridge the models' predictions with the real-world behaviors. We will further elaborate on each of these problem types in subsection 1.2.

Before moving on, we would note that detailed Reviews have been published in the literature on data-driven plasma science [18] as well as on ML promises for modelling and analysis of low-temperature plasmas (LTPs) [19]. Additionally, in Ref. [20], Bonzanini et al. provided an in-depth description of the ML fundamentals for LTP applications and discussed some specific use cases of ML to study LTP systems. A Perspective article has also been published regarding the applications of ML to plasma medicine [21].

Inspired by the developments in the landscape of scientific machine learning over the past few years, particularly the advances in ML for computational fluid dynamics (CFD), the present Perspective complements the above publications. Specifically, given the analogous challenges in both fields, our paper aims to sketch a roadmap for how the progress in ML for CFD, as exemplified by several notable efforts, can be transferred to computational plasma physics. The paper discusses new viewpoints, specific pathways and research directions to realize the full benefits of ML for plasma modelling with the aid of specific demonstrative examples.

The prior Review and Perspective papers, and the references therein, can be referred to for a more extensive reading beyond and in addition to the scope of this work.

### 1.1. Overview of the various categories of ML algorithms

We describe below each of the learning categories (types of ML algorithms) enumerated in the preceding Section and provide an overview of the most common algorithms belonging to each category.

(1) **Supervised Learning**: In supervised learning, corrective information are available to the learning machine. Simply put, the algorithm is trained on a dataset consisting of input-output pairs, where each input is associated with a corresponding output or label. The goal is to learn a mapping or function that predicts the output for new, unseen inputs based on their comprising features. The parameters of the learning machine are determined through the process of minimizing a defined loss function, which is commonly in the form of an L2-norm error between the model's prediction and the "ground-truth" [labeled] input data. Common architectures for supervised learning are briefly described below.

  i. **Linear and generalized linear regression**: Linear and generalized linear regression are foundational techniques in statistics as well as machine learning, providing interpretable models for understanding the relationship between variables and making predictions based on the observed data.

  Linear regression [22] is a widely used method for predicting a continuous output based on one or more input features. It assumes a linear relationship between the predictors (independent variables) and the target (dependent) variable. This method is particularly effective when the relationship between variables is or can be approximated as linear.

  Generalized linear regression [23] expands upon the linear regression by accommodating, e.g., nonlinear relationships between variables. It encompasses a broader class of models, where the target variable follows a distribution from the exponential family, for instance, Gaussian, binomial, Poisson. It enables modelling relationships between variables that may not be strictly linear. It can also handle various types of target variables, including binary outcomes and continuous variables with non-constant variance. Generalized linear regression provides a flexible framework for modelling diverse datasets, hence, serving as a versatile tool in data analysis and modelling.

  ii. **Support vector machines (SVM)**: SVMs [24] are versatile algorithms typically used for classification and regression tasks. SVM finds the optimal hyperplane that separates different classes or fits the data with a minimal margin of error. In the context of classification, SVMs sit adjacent to another widely used algorithm, the random forests [25]. SVMs have been one of the most successful supervisory data-mining algorithms in the last few decades [22]. The emergence and rapid advances of deep neural networks (see Point iii below), however, implies that when sufficient data is available, SVMs are superseded by the deep neural nets.

  iii. **Neural networks**: Neural networks (NNs) are computational architectures that are loosely inspired by the biological networks of neurons in a human brain. They serve as powerful nonlinear function approximators and are commonly used in supervised learning. They consist of interconnected neurons organized in layers, where each neuron processes inputs through an activation function to produce an output. The modular structure of neural networks allows for flexibility in combining neurons into various architectures that are tailored to different problem domains and data types. NNs are closely connected to the universal approximation theorem by Hornik et al. [26], which states that any function can be approximated by a sufficiently large and deep neural network. Deep neural networks (DNNs), which consist of several "hidden" layers of neurons between the input layer and the output layer, have shown remarkable capabilities, largely due to their ability for hierarchical learning. This means that whereas the initial layers of a DNN recover the simple and basic relationships in the data, deeper layers use a combination of the extracted information to derive more abstract relationships. This is highly significant and relevant for physical systems since many such systems also exhibit hierarchical behavior.

Feed-forward NNs are one of the most common architectures. Optimization methods such as back-propagation [13] adjust network weights to minimize the error between predictions and labeled training data. In a neural net, if the activation functions are comprised by convolutional kernels, the network is called a convolutional neural network (CNNs) – a powerful architecture which is renowned for image and pattern recognition tasks [27].

Recurrent neural networks (RNNs) are another type of NNs that are pivotal for the processing of sequential information, such as the time-series data. They retain memory of past inputs through recurrent connections, allowing them to capture temporal dependencies inherent in sequential data. However, traditional RNNs face challenges with vanishing or exploding gradients during training. The introduction of long short-term memory (LSTM) algorithms [28] addresses these issues by incorporating memory cells and gating mechanisms, enabling efficient transmission of long-term information. Extensions like multi-dimensional LSTM networks (MD-LSTM) [29] further enhance their capability to handle high-dimensional spatiotemporal data.

In addition to RNNs, alternative architectures like echo-state networks [30] (belonging to the broader category of "reservoir computing") have also emerged as potent tools for predicting dynamical systems.

(2) **Unsupervised Learning**: Here, the learning task is to extract patterns and structures from unlabeled data without explicit supervision in order to discover hidden patterns, groupings, or clusters within the data. The typical problem types encountered in unsupervised learning include clustering and dimensionality reduction. Some common architecture are:

  i. **K-means clustering**: This is the most common and widely used algorithm for clustering, i.e., the task of dividing data into different groups based on similarity or distance metrics. K-means clustering partitions the data into "k" clusters, resulting in the transformation of the data space into several Voronoi cells [16].

  ii. **Proper orthogonal decomposition (POD)**: POD, also referred to as linear principal component analysis (PCA), is dimensionality-reduction technique that projects high-dimensional data onto a lower-dimensional subspace while preserving the maximum variance in the data. POD/PCA algorithm has been playing a central role in reduced-order modelling for decades. This is because POD enables deriving the dominant modes underlying data from a system. The underpinning computing method for POD/PCA is the Singular Value Decomposition (SVD). To link POD, the linear PCA or SVD to machine learning, we can imagine these algorithms as a shallow neural net with linear activation functions. We will elaborate on this connection and discuss POD/SVD further in subsection 4.1 in the context of dimensionality reduction for reduced-order modelling.

  iii. **Autoencoders**: These are NN architectures designed to learn efficient, compressed representations of the input data by first encoding it into a lower-dimensional space (latent space) and then decoding it back to the original input. Autoencoders are used for tasks such as feature extraction and denoising. The autoencoder architecture is a nonlinear generalization of the POD/SVD method, achieved by extending the number of layers and incorporating nonlinear activation functions. A closely related architecture is variational autoencoder [31], which can be considered as generative ML because it produces new outputs from the reduced-dimension latent space representation of the input data.

  The power of autoencoders to learn nonlinear manifold coordinates to represent the data can enable achieving improved compression in the latent space [32]. As such, autoencoders can play a crucial role in the development of reduced-order models (ROMs), and, hence, we will have a closer look at these architectures as well in subsection 4.1.

(3) **Semi-supervised Learning**: The algorithms belonging to semi-supervised learning learn under partial supervision, meaning that either the amount of labeled data is limited, or the corrective information are received directly from the learning environment. Generative adversarial networks (GANs) [33] and Reinforcement Learning (RL) algorithms [34] are the main semi-supervised learning algorithms.

  i. **Generative adversarial networks (GANs)**: GANs are learning algorithms designed to produce generative models capable of generating data that closely resembles the distribution of the training data. The essence of GANs lies in a dynamic interplay between two neural networks: the generative network and the discriminative network. In this setup, the generative network generates candidate data samples, while the discriminative network evaluates these samples to distinguish between real and synthetic data. The training process resembles a zero-sum game [33], where the generative network aims to produce data examples that deceive the discriminative network into misclassifying them as real data. Conversely, the discriminative network tries to accurately differentiate between real and generated data. Through this adversarial learning process inspired by game theory, the networks iteratively refine their parameters to improve their performance.

The adversarial training paradigm inherent to GANs allows the networks to learn from each other without the need for explicit supervision, leading to a so-called "self-supervised" learning process. While this training approach adds to the interest and appeal of the GANs, there remains a challenge in ensuring convergence to an equilibrium point in the adversarial game. Although larger datasets can enhance the training process, convergence is not guaranteed [16]. This highlights an area for further research on GANs.

ii. **Reinforcement Learning (RL)**: RL is a mathematical framework [34] designed for solving complex problems through goal-directed interactions between an agent and its environment. Unlike supervised learning, where the agent learns from labelled information, RL operates in a semi-supervised manner, where the agent learns from its own experiences, guided by infrequent and partial rewards. The primary objective of RL is not merely to uncover patterns in actions or in the environment but to maximize long-term rewards. In RL, two fundamental components are the agent's policy and the value function. Agent's policy defines the mapping between states of the system and optimal actions. The value function assesses the utility of reaching a particular state for maximizing long-term rewards.

RL poses significant computational challenges due to the large number of episodes required to effectively account for the agent-environment interaction. Additionally, a core challenge in RL is the long-term credit assignment problem, especially when rewards are sparse or delayed over time [16]. This challenge involves inferring causal relations between individual decisions and rewards from a long sequence of states and actions. Efforts to address these issues include augmenting sparsely rewarded objectives with densely rewarded subgoals [35], aiming to enhance the learning process and facilitate effective decision-making in complex environments.

## 1.2. Primary areas for ML applications to physical systems

Within the introductory discussions presented before, we divided the applications of ML for numerical modelling into two main categories or problem types: **(1)** capturing expensive physics cost-efficiently, and **(2)** discovery of the system dynamics and discrepancy modelling.

For the first problem type, we can use ML to either accelerate a computationally costly high-fidelity simulation (Section 2) or to develop a data-driven surrogate for the high-fidelity simulation using its output data (Section 4). Such a surrogate enables low-computational-cost (fast) predictions and analyses and is often also referred to as a reduced-order model (ROM).

The second problem type actually comprises two correlated sub-problems: dynamics discovery, and discrepancy modelling. In dynamics (model) discovery, the aim is to discover new physics and/or new physical correlations/relationships, for which governing equations may either not exist or that may be difficult to develop using first-principles. For example, we can imagine discovering models from data for phenomena in epidemiology, neuroscience, or finance, where we do not have a pre-defined set of governing equations.

The other sub-problem involves complementing an existing system of equations for a physical system by finding relations and/or models for the physics that are not captured, hence, reconciling the simulations and the real-world behaviors. In this discrepancy modelling application [36], we may, e.g., aim to find closure terms for the system of equations or intend to augment the system of equations by enabling it to capture multiphysics interactions. As an example, in fluid-based modelling of plasma systems, we already have a partial knowledge of the set of equations that describe the systems' dynamics. Thus, we can apply ML to find the missing closure terms/models for the plasma fluid equations, hence, addressing their currently lacking self-consistency.

The two general problem types described above are not mutually exclusive. Rather, they share certain commonalities in terms of the objectives, as well as the algorithms employed. For instance, dynamics discovery (problem type 2) is also an essential component of reduced-order modelling (problem type 1). In fact, in the ROM development context, the aim is often to model the time evolution of our physical system in a new, typically reduced-dimensional coordinate system. In such a coordinate, the governing equations are not necessarily known, hence, the need for the discovery of dynamics in the identified reduced coordinate. We will elaborate on this aspect later in Section 4.

Selection of the problem type that we intend to solve using ML affects consequently the data that needs to be curated as well as the choice of the ML algorithm from the general categories of architectures reviewed in subsection 1.1.

Two aspects closely linked to the ML architecture selection are the definition of a loss function (for instance, an L2-norm error between the predictions and the ground-truth for supervised learning) and the choice of an optimization algorithm. The optimization acts to minimize the defined the loss function during the architecture's training by adjusting the parameters of the ML model.

There are additional points worth highlighting here: first, in the recently emerged framework of "physics-informed" ML [37], the primary objective is to embed prior knowledge of physics into the learning machine so as to achieve higher levels of generalizability, robustness, and interpretability. Conservations, symmetries, and invariances constitute various ways by which physics can be incorporated into ML.

Second, the concept of model "parsimony" is another approach for physics embedding. This concept is inspired by the observation that all generalizable and interpretable physics laws we currently have (Newton's second law or Maxwell's equations, e.g.) feature as few terms as necessary to describe the relevant phenomena. We will revisit this important idea and discuss this further in Section 4.

The incorporation of the physics into ML can be manifested within the loss function definition and/or within the optimization algorithm. The former embedding acts to "promote" physics because the physics-related loss term(s) would be competing against other loss factors, for instance, the L2-norm error. On the contrary, physics embedding within the optimization algorithm serves to "enforce" physics in the learning process of the machine. Embedding physics within the optimization procedure is, nonetheless, usually effort-intensive.

When considering physical knowledge embedding within the loss function of an architecture, a well-known and prevalent family of ML algorithms are physics-informed neural networks (PINNs) [16][38][39]. Pioneered by Raissi et al. [38], PINNs offer an elegant approach to solving partial differential equations (PDEs) by incorporating the satisfaction of the underlying physics of the problem directly into the loss function of a neural network. This method leverages deep learning principles to tackle complex PDEs, utilizing automatic differentiation within the back-propagation algorithm to compute partial derivatives and formulate the equations to be enforced through the loss function. PINNs have been used for a wide variety of applications, including turbulence modeling [40], developing ROMs [41], and accelerating traditional solvers [42]. PINNs have shown notable capability in handling noisy data [43], enabling robust predictions in scenarios where data uncertainty is present.

Nonetheless, one notable challenge with PINNs is their underlying assumption that the system of equations describing a system's evolution is fully known so that satisfying the equations can in fact be included within the loss function of the architecture [44]. Despite this, PINNs can be effective for discrepancy (closure) modelling in case the problem can be formulated as an inverse problem or system identification, which PINNs are able to handle [44].

Informed by the above introductory materials and discussions, in the remainder of this Perspective article, we examine promising research directions for ML in computational modelling along three distinct paths: **(1)** accelerating high-fidelity, computationally heavy simulations, **(2)** discrepancy modelling and closure model development, and **(3)** reduced-order modelling.

We discuss these pathways in Sections 2 to 4. We review some of the most relevant and notable prior research, from fluid mechanics, and also from plasma physics where available. By spotlighting the promising developments in ML for fluid mechanics, we aim to paint an overall picture of the opportunities for cross-disciplinary transfer of ideas and advances into computational plasma physics.

Section 5 provides the concluding remarks. These mainly surround the challenges on the path of using ML for modelling complex physical systems, including plasmas, as well as the prime applied areas where ML-enabled plasma modelling can be of great value.

**Section 2: Accelerating high-fidelity simulations**

In order to set the context, and to provide an overall view of various possibilities regarding the use of ML for accelerating high-fidelity simulations, we first discuss a number of interesting examples from CFD. Afterwards, we review some prior efforts performed in plasma physics to accelerate/enhance simulations using ML.

Many efforts in fluid dynamics have focused on improving the cost-efficiency of direct numerical simulations (DNS). DNS is the highest fidelity modelling approach in CFD, which aims to solve the Navier-Stokes system of equations for the fluid flow across all relevant spatiotemporal scales, from the smallest turbulence scale to the largest and, hence, requires very fine-resolution grids. The computational cost of a DNS simulation roughly scales with the third power of the Reynolds number [45]. Thus, the method is computationally burdensome even for moderate values of Reynolds number. Accordingly, researchers have aimed to accelerate these simulations using ML along two main paths: **(i)** improving the discretization schemes, and **(ii)** developing correlations to link fine- and coarse-grid simulations.

Before looking at examples of prior work along each path, it is important to underline one point. For plasmas, it is often difficult for a fluid system of equations to properly describe the behaviors and the dynamics. In many applications, plasmas are in thermodynamic nonequilibrium, exhibit strong anisotropies, and their global response is majorly affected by the underlying microscopic instabilities and/or turbulence. As a result, with some exceptions such as the case of ideal magnetohydrodynamics (MHD) [46], a system of equations for plasmas that may (partly) resemble Navier-Stokes will not be able to generally describe a wide range of plasma conditions. In fact, plasma fluid systems of equations for most practical applications and scenarios lack closure models to represent the missing kinetic effects (moment closures) and the nonclassical effects, such as the influence of instabilities (transport closures) [47][48].

In any case, methods to accelerate DNS using ML may still be applied in plasma physics modelling from different perspectives. In particular, MHD simulations, where applicable, and other high-fidelity grid-based models, such as direct-kinetic simulations that aim to solve the Vlasov-Maxwell equations in a Eulerian manner [49], can benefit from ML-enhanced discretization schemes as well as ML-enabled correlations to map the results of a low-resolution simulation to a high-resolution one.

With the above remarks in mind, we now review some notable examples of accelerating high-fidelity simulations using ML approaches in fluid dynamics. In this regard, toward improving the discretization schemes, the method of "stencil learning" was pursued by authors such as Bar-Sinai et al. [50] and Zhuang et al. [51], who used deep learning to estimate the spatial derivatives such that the learned numerical discretization retains high accuracy even with low-resolution computation grids. Stevens and Colonius [52][53] and Jeon et al. [54] worked to enhance the accuracy of the classic finite-difference and finite-volume schemes using ML and deep learning. In particular, Stevens and Colonius proposed the FiniteNet [53] – an architecture based on Convolutional and LSTM neural networks – to improve solving time-dependent PDEs.

Toward developing correlations to link fine- and coarse-grid simulations, a prominent work has been carried out by Kochkov et al. [55], who pursued the ideas of super-resolution from image processing to derive a deep-learning-based correction that enables resolving detailed features of a turbulent flow evolution with a coarse-grid simulation (Figure 1). Their method enabled capturing the results of a high-resolution reference simulation with mesh resolutions that were 8-10 times coarser along each dimension [55].

A similar work has been performed by Fukami et al. [56], who employed CNNs to reconstruct high-resolution turbulent flow fields from highly under-resolved field data. In addition, there have been other efforts to improve the performance of PDE solvers on coarse grids, e.g., by Li et al. [57].

ML has been employed as well to solve Poisson-like equations, which are encountered in fluid dynamics, and are evidently of high relevance in plasma physics. For example, Tang et al. [58] have assessed developing deep-learning-based Poisson solvers, and Ajuria et al. [59] have applied deep learning to solve the Poisson-type corrector step in the incompressible flow simulations.

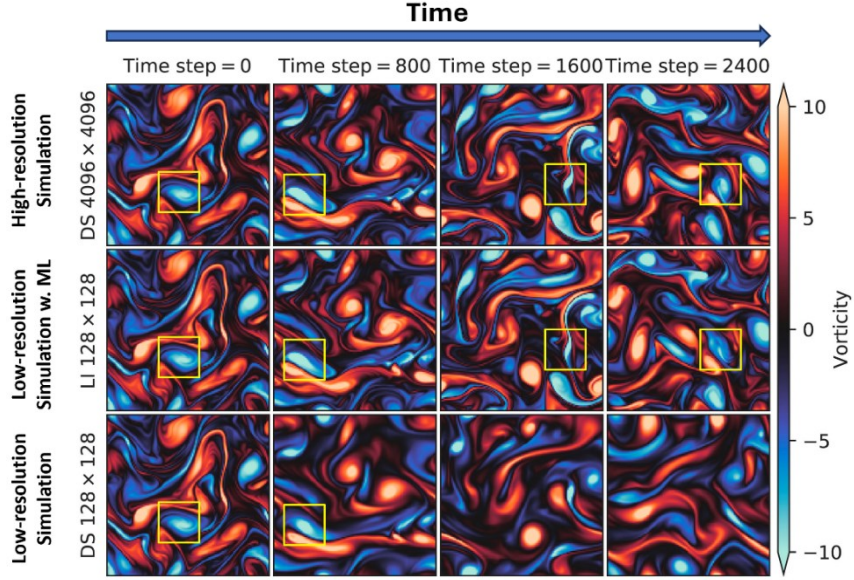

Figure 1: Sample figure from the work of Kochkov et al. [55] (super-resolution acceleration of numerical simulations). The plot shows the time evolution of the vorticity field from simulations with a fine (**top row**) and a coarse grid (**bottom row**) as well as a coarse-grid simulation enhanced with ML (**middle row**). The vorticity field evolution is shown at four different time instants, and some key vortical structures in the flow are specified by yellow square boxes. Reproduced from Ref. [55], available under CC-BY NC-ND license.

We would note at this point that super-resolution and subgrid modelling techniques are typically applicable for the types of grid-based simulations in which the size of the computation cells can be increased without affecting the stability of the simulation. In such scenarios, some details of the physics and/or the phenomena (like turbulence) that occur at spatial scales below the resolution of the computational mesh may be lost, but the simulation as a whole will not become unstable.

From another perspective, in the fluid simulation examples such as the work of Kochkov et al. [55], resolving the turbulent scales below resolution of the simulation did not affect the global behavior of the flow (i.e., absence of inverse energy cascade behavior [60][61]).

It naturally follows from the above statements that conventional kinetic PIC simulations for plasmas do not provide any evident pathway to use the super-resolution ideas. This is due to stringent stability requirements that exist on the spatiotemporal resolutions of the traditional (explicit) PIC simulations [5][6]. Resolving the smallest scale phenomena is in fact essential in many plasma systems in order to obtain the correct global behavior and steady state.

Nevertheless, the need for cost-efficient high-fidelity plasma simulations – some reasons for which were elaborated on within Section 1 – has led to other ways by which the researchers have tried to employ ML to accelerate PIC simulations.

Toward this purpose, ML and data-driven techniques have often been employed to substitute a specific, computationally intensive module of PIC codes, e.g., the Poisson solver [62]. Aguilar and Markidis [63] employed deep learning (in the form of a fully connected neural network and a CNN architecture) to enhance a PIC simulation by learning a mapping between the electric field and the phase-space information of the particles. They assessed their DL-PIC algorithm against the two-stream instability test case, observing that the code provides good predictions in this case [63]. However, they reported that their DL-PIC does not conserve momentum and energy, requiring explicit enforcing of these in the DL field solver as the next step [63].

Kube et al. [64] proposed the use of neural networks to accelerate implicit, energy-conserving PIC algorithms (e.g., see Ref. [65]) by reducing the number of GMRES (generalized minimal residual) iterations required to advance the PIC solution in time. Vigot et al. [66] explored the application of Graph Neural Networks (GNNs) to solve the Poisson equation, with the aim of accelerating this often computationally heavy step in explicit PIC simulations [5][6].

Nicolini et al. [67] and later Nayak et al. [68] aimed at accelerating the field-solve step of the PIC simulations by substituting the high-dimensional field solver with reduced-order models for the electromagnetic fields. Nicolini et al. used POD for this purpose [67], whereas Nayak et al. employed the DMD method [68]. Despite promising results, the generalizability of the derived ROMs for electromagnetic fields in extrapolation, whether in time or in parameter space of a plasma system, was highlighted as a challenge that needs to be addressed [67]. In Ref. [69], Nayak et al. used the DMD method with a sliding window to identify the system state transition from transient to equilibrium, allowing the identification of the limit cycle, and hence predicting the electric field evolution and the resulting plasma particles dynamics. The authors discussed the implications of this approach for accelerating PIC simulations [69]. Badiali et al. [70] proposed the application of ML-based models to speed up the multiphysics modules of a PIC simulation, such as the collisions operator.

In a separate context to the above efforts but still with the aim of accelerating plasma simulations, Zhong et al. used PINNs in Ref. [71] to solve stationary and time-dependent PDEs governing thermal plasmas, particularly 1D arc discharges.

In Ref. [72], we proposed using ML-enabled super-resolution to further enhance the computational cost-efficiency of the reduced-order PIC scheme [73]-[75]. We discussed that, contrary to the conventional PIC simulations, the scheme of the reduced-order PIC and its underpinning formulation are readily amenable to super-resolution and subgrid modelling approaches. This is because of the innovative way that such a simulation decomposes a multidimensional problem in space using coarse grids while respecting the stability criteria of explicit, momentum conserving PIC implementations [72].

The proposed ML-enhanced reduced-order PIC involves a deep-learning (DL) module that augments the spatial resolution of the resolved electric field – the main temporally evolving driving force in electrostatic plasma applications – while interacting with the other modules of the PIC simulation. By enhancing the resolution of the electric field, the particles under the influence of the enhanced field can essentially experience the full-2D/3D effects in a reduced-order (or quasi-multidimensional) PIC simulation [76]. This application implies a physics-constrained approach to DL enhancement since the aim is not to replace the field solver with an ML surrogate, but rather to improve the resolution of the outputs of the field solver, which are obtained under the constraint of satisfying Gauss' law.

We presented in Ref. [72] an architecture called Shallow Recurrent Decoder (SHRED) [77][78] as a promising candidate for super-resolution purposes, including the application to the reduced-order PIC described above. We will discuss the architecture of SHRED in more detail within subsection 4.2.1.4 in the context of dynamics discovery for reduced-order modelling. Here, we present a sample result through Figure 2, showcasing the utility of the architecture for mapping low-resolution snapshots of a plasma property to accurate high-resolution spatial reconstructions.

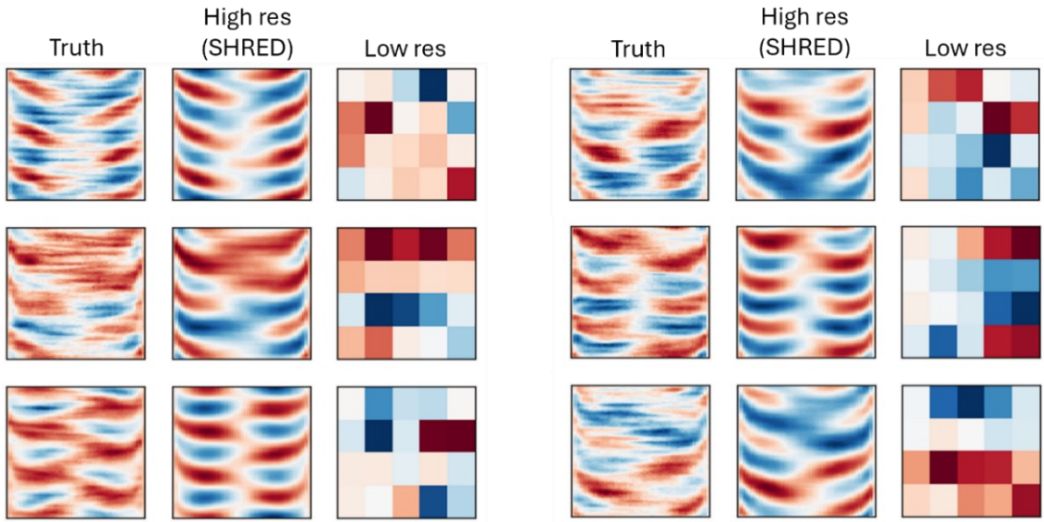

Figure 2: Demonstration of SHRED's capability for super-resolution enhancement of low-resolution data. The data are obtained by applying a convolutional-pooling averaging filter to the output snapshots of a high-fidelity 2D PIC plasma simulation. The PIC simulation was performed in a 2D configuration representative of a radial-azimuthal cross-section of a Hall thruster – a plasma technology for spacecraft propulsion. The various subplots show normalized 2D snapshots of a plasma property – electron's axial current density ($J_{ey}$) – at six random time instants within a test window. **Truth** snapshots are from the reduced-order quasi-2D PIC simulation, **low-res** snapshots are the spatially down-sampled dataset, and the **high-res** snapshots are the super-resolution-enhanced dataset using SHRED.

**Section 3: Discrepancy modelling and closure model development**

Discrepancy modelling represents an exciting and important frontier in scientific machine learning. It is in fact one of the most promising ways by which ML can be integrated in computational modelling and engineering, bridging the simulated and the physical world. Discrepancy modelling is closely related to data assimilation, particularly from the perspective of model error quantification and correction.

Discrepancy modelling is, nonetheless, a more general framework, comprising a broader set of approaches to address errors and/or inaccuracies in traditional (first-principles) physics-based models used for simulations and predictions. These errors can, in part, arise from missing physics, which, in turn, can significantly affect the predictions of complex physical systems using the first-principles models. ML techniques have, thus, been developed to learn models for these discrepancies. For example, Levine and Stuart [79] recently proposed leveraging recurrent neural networks (RNNs) to model hybrid problems in non-Markovian settings, demonstrating the potential of ML in capturing missing physics and, hence, improving dynamical models.

By augmenting known first-principles models with discrepancy models learned from data, it is possible to improve the accuracy and reliability of dynamical models (refer to Figure 3 and the description in the figure's caption). Historically, early models of systems were often coarse approximations of the underlying physics, leading to discrepancies between the model and the reality. While computational advancements have enabled more detailed simulations (models), model-measurement mismatch has remained a challenge across various fields. This challenge has been especially pressing for plasma systems due to the complexities of the physics in realistic settings, on the one hand, and the computational burden and/or incompleteness of the plasma simulation tools and models, on the other.

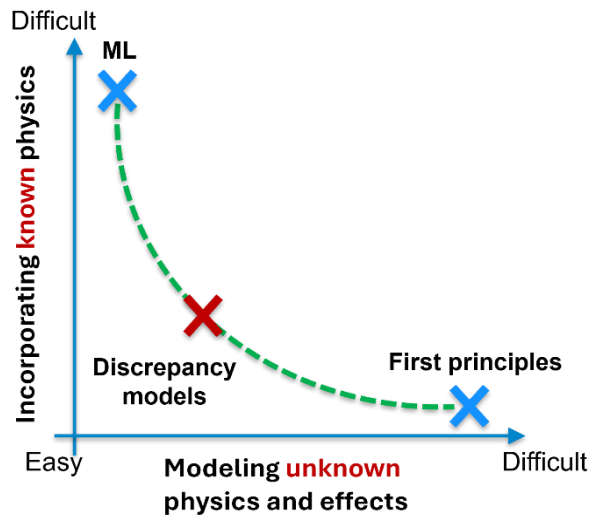

Figure 3: Schematic diagram of discrepancy modelling as an alternative path to inform ML-based dynamics discovery with the known physics. Discrepancy modelling in this sense facilitates learning unknown physics and effects that may be difficult to model via first-principles methods.

In modern applications, discrepancy modelling has become increasingly relevant, especially for use cases such as digital twins, where precision is crucial. Predictive and scalable digital twin technologies will rely on computational models [4][80]. However, if derived from first-principles physics solely, these underpinning models will at least in plasma physics fail to accurately match reality. This is where discrepancy modelling becomes crucial because it provides a systematic way for improving the models with data in order to ensure that they align increasingly better with real-world observations.

ML techniques, such as neural networks, have emerged as powerful tools for learning discrepancy models directly from data, providing a data-driven framework to address model inaccuracies and enhance predictive capabilities. In this respect, whereas Levine and Stuart [79] derived discrepancy models using NNs in a dynamical representation, e.g., in the form of differential equations, Ebers et al. [36] considered a broader class of ML/data-driven techniques and also explored learning a discrepancy model in the state space (i.e., addressing the residual errors in the approximation of a system state). Their work demonstrated the versatility and potential of different DD approaches in capturing and modelling discrepancies in physical systems [36].

Developing closure models for the system of governing equations can be considered as a subcategory of discrepancy modelling. Data-driven closure finding represents significant opportunities for plasma systems. The closure problem is an active subject of research in computational plasma physics. Several approaches have been pursued so far in this regard [81]-[86], but achieving generalizable and self-consistent closure models for the plasma fluid equations has remained elusive, nonetheless. This is while plasma fluid descriptions are of interest due to their higher computationally affordability compared to kinetic simulation approaches, such as PIC.

In the most general terms, the closure problem in plasma physics refers to two aspects: first, models that can represent higher-order moments of the particles' distribution function, capturing deviations from an isotropic, Maxwellian distribution function [87][88], in terms of the lower-order moments (density, mean velocity, and pressure tensor); this is referred to as "moment closure". The second aspect of the closure problem is to develop models that resolve the nonclassical effects of plasma phenomena, such as instabilities and turbulence, on the momentum and energy transport of plasma particles; this aspect of closure can be called "transport closure".

In fluid mechanics, when the Navier-Stokes equations are valid (i.e., the case of continuum fluids in thermal equilibrium), the closure problem is often synonymous to finding a turbulence model that obviates the need for highly fine-resolution DNS simulations. In this respect, turbulence models for the Reynolds-Averaged Navier-Stokes (RANS) equations try to correlate the so-called "Reynolds stress tensors" to the mean flow properties. Turbulence closure models for the Large Eddy Simulations (LES) aim to resolve the effects of the flow turbulence that cannot be captured because they fall below the grid resolution. In any case, moment closure problem can also arise in fluid mechanics, for instance, in case of the rarified and/or non-equilibrium gases, for which the Navier-Stokes equations can become less valid [89].

Accordingly, despite certain differences in what constitutes a closure problem in fluids vs plasmas modelling, the ML applications toward closure finding in fluid dynamics can still be relevant and beneficial in plasma physics domain. In this regard, we review below two specific examples of prior work in fluid mechanics. A more detailed review of ML for closure modelling of fluids is provided by Duraisamy et al. in Ref. [90].

The first example showcases how incorporation of partial knowledge of the physics improves the accuracy and performance of the ML-derived closure models. Ling et al. [91] developed a deep neural network architecture that featured a multiplicative layer with an invariant tensor basis in conjunction with the last hidden layer of the network before a merged output layer [91]. The inclusion of this multiplicative layer enforced the closure model learned for the Reynolds stress tensor to respect Galilean invariance, meaning the model was valid in any inertial frame of reference. By embedding this physics into the network, the authors showed an improved performance of their RANS closure model, better than any traditional model, even for the cases that are challenging to predict with the RANS models [91].

As for the second example, Novati et al. [92] followed an agent-based approach (Reinforcement Learning) to estimate and determine more accurately the free parameter (coefficient) of a subgrid scale (SGS) model in a coarse-grid LES simulation. Several RL agents were deployed at various fixed grid locations within the simulation domain. These agents computed the values of the SGS model's coefficient at their grid locations over time and learned under the influence of a policy and a reward. The policy depended on the state of the agents that incorporated both local and global variables [92]. The reward, which was also dependent on the local and global variables, was received based on the accuracy of the simulation [92]. The RL-based approach pursued by Novati et al. [92] exhibited good generalization characteristics across different grid sizes and flow conditions. The outcomes from this work open interesting prospects for closure finding and subgrid modelling, as appropriate, in plasma simulations' context.

In plasma physics, discrepancy modeling has recently gained traction in the community, particularly in the areas of data-driven closure modeling and data assimilation.

In the area of closure model development, Donaghy and Germaschewski [93] employed a sparse regression algorithm – Sparse Identification of Nonlinear Dynamics (SINDy) [94] (see subsection 4.2.1.3) – to derive from kinetic PIC simulation data of a magnetic reconnection configuration – the "*collisionless Harris sheet reconnection problem*" – a ten-moment fluid system of equations for the plasma with a closure model for the tenth moment term, aiming to improve the "*Hammett–Perkins closure*". The salient finding was that even though the closure term from SINDy was improved over the Hammett–Perkins closure for the nonlinear phase of the system evolution, it was not adequate for the linear phase [93], hence, warranting further work along the direction of closure discovery with SINDy.

Further to the application of ML for closure model discovery, van de Plassche and collaborators [95] developed a fast feed-forward neural network, QLKNN, to find ML closure (surrogate) models for the heat and particle transport fluxes from a tokamak's core from a database comprising 300 million flux calculations of "*the quasilinear gyrokinetic transport model QuaLiKiz.*" Integrating these ML models into a tokamak modelling framework and a control-oriented tokamak transport solver, they reported orders of magnitude speed-up while maintaining a good accuracy [95]. Heinonen and Diamond in Ref. [96] applied a deep neural network to data from a direct numerical simulation of the 2D Hasegawa-Wakatani plasma system in order to derive mean-flow closure terms for the turbulent transport fluxes, showing promising results.

In the context of technological low-temperature plasmas, Jorns [83] focused on complementing fluid models of the discharge in a Hall thruster – a prominent in-space plasma propulsion technology. He followed symbolic regression (see subsection 4.2.1.3) to find a functional relationship for the "anomalous" electron collision frequency in terms of the local plasma properties [83]. The derived data-driven symbolic function showed a predictive capability as well as a quantitative improvement over five first-principles models of the anomalous collision frequency [83]. The limitations of the approach, partly in terms of the extensibility of the derived functional forms to parameter spaces notably outside the training range, were examined and discussed [83].

In the area of data assimilation, some interesting efforts have been made over the past few years. Greve et al. [97] presented a data-driven approach for model calibration utilizing the Wasserstein metric so that the parameters of a dynamics model, such as one for a Hall thruster discharge, can be optimized to a reference solution, which may come from high-fidelity simulations or experiments. Another effort by Shashkov et al. [98] pursued a similar general objective, using Bayesian optimization with a Gaussian process to calibrate the electron conductivity (transport) profile in a Hall thruster model to match the experimental data from the thruster.

A study by Eckhardt et al. [99] presented the shadow manifold interpolation (SMI) technique, a nonlinear method for mapping and reconstructing spatiotemporal dynamics in Hall thrusters. They demonstrated that this technique outperforms the traditional approaches based on fast Fourier transform (FFT) by accurately reconstructing signals even in the presence of noise and non-periodic data, despite being more computationally intensive and facing challenges such as finding the optimal lag [99].

Greve and Hara in Ref. [100] and Hara in Ref. [101] evaluated the application of different variants of Kalman filters, including the extended and the ensemble variants, to the problem of data assimilation in various plasma systems. These works demonstrated the utility of these filtering methods in estimating and inferring parameters and properties that are challenging to access (measure), by combining physics-based models with experimental data.

While the above efforts focused on low-temperature and technological plasmas, data assimilation has also been used recently within the fusion plasma domain. For instance, Sanpei et al. [102] presented an efficient MHD equilibrium reconstruction method using an adjoint-based algorithm. The proposed method utilizes data assimilation to refine model predictions with real-time data from sensors, and while it has been specifically designed for axisymmetric toroidal plasmas, the authors argued its extensibility to other plasma configurations [102]. Morishita et al. [103] demonstrated a data assimilation approach to enhance the adaptive control of magnetic fusion plasmas. Tested on the Large Helical Device (LHD) in Japan, predictive control of plasma electron temperature by adjusting heating power in real-time based on observations was demonstrated [103].

**Section 4: Reduced-order modelling**

Reduced-order modelling is a powerful technique, which is used to simplify the descriptions of complex systems by representing their essential dynamics in a lower-dimensional space, typically by focusing on dominant coherent structures underlying the system's evolution.

To aid further elaborations on reduced-order modelling and to set certain perspectives, we have presented in Figure 4 a graph of how various possible plasma models can be spread across the spectrum of computational cost vs accuracy and generalizability.

A ROM is defined in reference to a full-order model (FOM) and provides a simplified representation of it, which retains essential dynamics of the original system while significantly reducing computational complexity. This simplification is achieved by reducing the number of degrees of freedom (dimensionality), variables, or equations describing the system. In the context of plasma physics, the systems generally evolve within a six-dimensional phase space and over time. Consequently, as Figure 4 depicts, the reference FOMs in plasma physics can be kinetic simulations. These models offer the highest fidelity level but also have the highest computational cost.

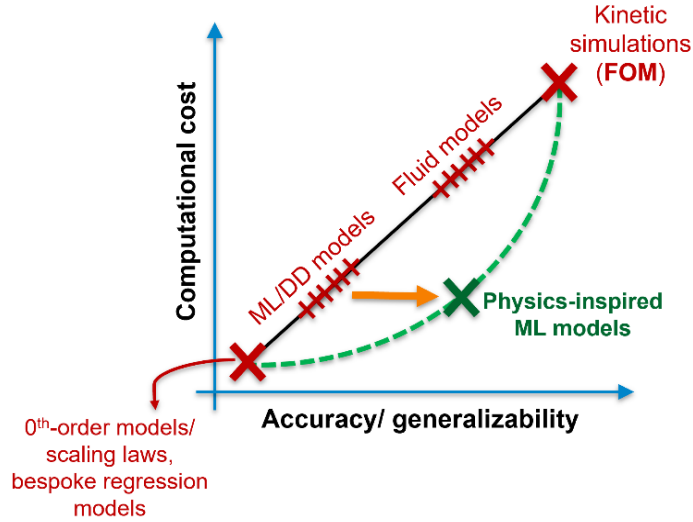

Figure 4: Graph of the computational cost vs accuracy and generalizability for various possible plasma physics models. The kinetic simulations represent full-order models (FOMs), whereas fluid models and ML/DD models are reduced-order models (ROMs). Incorporation of physics-inspired constraints into learning promotes accuracy and generalizability while maintaining computational cost at the level of ML/DD models purely based on data.

Noting the high dimensionality of the spatiotemporal space for plasmas' evolution, if one succeeds in establishing a closed 3D system of partial differential equations for a given plasma system, such system of PDEs can still be interpreted as a ROM. With this statement, plasma fluid models are a kind of reduced-order model because, to derive these models, we take velocity moments [87] of the plasma kinetic equation [104]. This procedure reduces the dimensionality of the problem in the most general case from 3D3V (three dimensional in configuration (physical) space and three dimensional in velocity (kinetic) space) to 3D. However, solving a 3D system of PDEs still entails significant computational cost, hence, placing fluid models at a rather intermediate position on the computational cost-accuracy plane.

It is noteworthy that zeroth-order models, scaling laws, and bespoke regression models are all some form of ROM as well. These models provide fast and computationally inexpensive solutions but lack the accuracy of more complex models and are hardly generalizable beyond the perimeters of their simplifying assumptions, problem descriptions, and/or the data for which they are developed.

ML and DD models fall somewhere between the extremes of accuracy/generalizability and computational cost. These models are faster alternatives to more expensive simulations and can, thus, enable rapid predictions for tasks such as design optimization and control. Traditionally, the high computational efficiency of purely data-based ML/DD ROMs comes at the expense of their generalizability. These models have usually been effective only for the specific geometries or parameter regimes on which they were trained.

We would point out that there are also classical non-ML-based approaches toward reduced-order modelling. One widely employed approach is the "POD-Galerkin" method. POD learns a low-dimensional coordinate system from data, enabling the representation of the dominant structures. A dynamical system is obtained within the POD-identified reduced subspace via performing a Galerkin projection of the governing equations (the Navier-Stokes equations for fluids, e.g.), onto these modes. While the POD step is data-driven, Galerkin projection requires a functional numerical implementation of the full governing equations [32]. In general, classical projection-based reduced-order modelling methods necessitate custom modifications to the numerical solver, making the approaches rather "intrusive" [32].

To address the challenges of ML-based ROM development, particularly with regard to the models' generalizability, we argue that alternative pathways for reduced-order modelling must be explored. The primary objective would be to maintain the computational cost at the level of ML/DD models based purely on data while enhancing the accuracy and generalizability of the models.

Toward this end, research has demonstrated that embedding prior or partial knowledge of the involved physics can significantly enhance the performance and generalizability of the ML/DD ROMs while also reducing the amount of training data required [32]. As was pointed out in subsection 1.2, this approach is often referred to as "physics-informed" [37] or, more broadly, "physics-consistent" ML modeling [16]. Examples of physics-based constraints include conservation laws, invariances, and symmetries, which can be incorporated as regularization terms in the loss function of the ML algorithm to penalize deviations from known physical behaviors (refer to subsection 1.2 as well).

The approach that we would propose is a nuanced version of the above. We refer to this approach as "physics-inspired" ML. Here, rather than only and strictly incorporating physical principles, the aim is to ensure that the model respects two key constraints: (**i**) low dimensionality, and (**ii**) sparsity (parsimony).

Concerning low dimensionality, one of the primary ideas in reduced-order modelling with ML is that, even within very high-dimensional complex systems, there are underlying dominant patterns that are crucial. For instance, in a megapixel snapshot of a simulation, not all individual pixels are important – only the dominant coherent structures matter. The number of these coherent modes is most often much lower than the total dimensionality of the system in the physical space (number of pixels, e.g.).

The concept of sparsity or parsimony is inspired by the observation that, in reality, dynamical systems can be represented by a few key terms and variables. Indeed, all fundamental physical laws and equations currently known to humanity are inherently sparse.

The general philosophy behind the framework of physics-inspired ML is to develop models that do more than merely fitting the data. While numerous ML algorithms can accurately fit the data, these models are often regarded as black boxes, i.e., they lack interpretability and generalizability beyond their training envelope. We aim for dynamical system models that are sparse and possess as few degrees of freedom as necessary to describe the behavior of the system.

Moving on, we would now look into the two main steps that are, generally speaking, associated with ROM development: first, finding a suitable reduced coordinate system (set of dominant modes) and, second, discovering a dynamical system model (a set of ordinary differential equations (ODEs) for instance) in the identified reduced coordinates.

We provide a clarifying example of these two steps: from high-fidelity, high-dimensional simulation data, a dimensionality-reduction technique, such as POD, can extract the dominant modes (structures) underlying the data. Then, a dynamics model can be constructed that describes how the amplitudes of these dominant modes evolve over time. Such a model is reduced order because instead of needing to resolve a large number of degrees of freedom, it now just resolves the time evolution of the amplitudes of a few spatial coherent structures.

We note that the above two steps for reduced-order modelling can be carried out consecutively or simultaneously. In case we aim to learn the dynamics asynchronously to finding a low-dimensional representation of our data, it may happen that the system would not have a sparse dynamical description in the identified reduced coordinate Hence, the simultaneous discovery of the reduced coordinates and the dynamics offers the advantage of constraining the learning algorithm to identify a coordinate system in which it will be possible to describe the dynamics in a parsimonious and, hopefully, generalizable manner.

Coordinate transformation and, rather equivalently, dimensionality reduction are crucial elements for developing predictive ROMs that hold the highest potential for generalizability and interpretability. To illustrate the importance of the concepts of coordinate transformation and dimensionality reduction, we consider two specific examples, which are visualized through Figure 5 and Figure 6.

Referring first to Figure 5, we compare the geocentric and the heliocentric views of the solar system. In the "naïve" geocentric coordinate system (Figure 5(a)), where Earth is placed at the center, the resulting dynamics appear complex, and no obvious laws describe the planetary motions in a straightforward way. However, when the coordinate system is transformed to place the Sun at the center (the heliocentric view, Figure 5(b)), the dynamical system becomes much simpler. Within this new framework, it is possible to derive simple, sparse models – Kepler's laws – that accurately describe planetary motion.

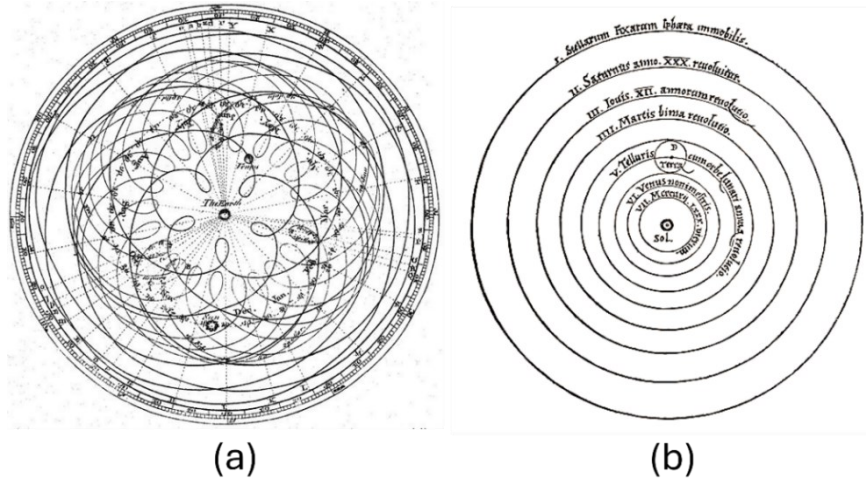

Figure 5: Visual demonstration of the concept of coordinate transformation for interpretable and generalizable physics modeling; models of the planetary motion: (a) the Geocentric Ptolemy's model, (b) the Heliocentric Kepler's model. Illustrations, available on the public domain, are reproduced from the webpage: https://www.space.fm/astronomy/planetarysystems/geocentricheliocentric.html.

Looking now at Figure 6, we imagine having a high-resolution video of a swinging pendulum. An ideal data-driven reduced-order model should be able to determine from the video that the minimum variable required to describe the motion of the pendulum is the angular displacement $\theta$, and subsequently discover the governing equation for the evolution of $\theta$ without any prior knowledge of the underlying dynamical system. However, learning both the coordinates and the dynamics from such high-dimensional data (i.e., the video) is not trivial. Nonetheless, by promoting sparsity and constraining the model to be as low-dimensional as possible, we can achieve this goal.

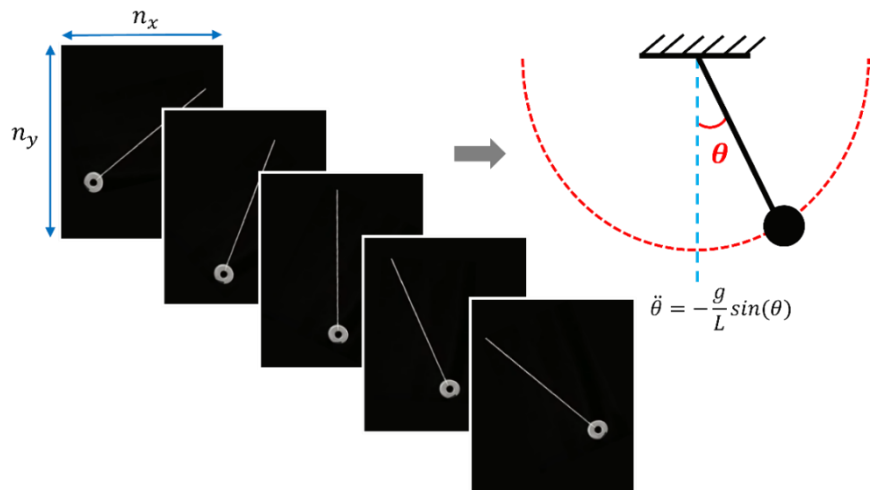

Figure 6: Visual demonstration of the role of dimensionality reduction to describe the underling dynamics; example of a swinging pendulum as seen from high-dimensional video snapshots of the motion vs as described by one essential variable, the displacement angle ($\theta$).

The above examples underscore the importance of coordinate transformation and dimensionality reduction on the path to developing effective and generalizable ROMs. By identifying the appropriate low-dimensional coordinate system, we can simplify the underlying dynamics, making it feasible to derive sparse, interpretable models that capture the essential behavior of complex systems.

In the following subsections, we focus on the two main steps for reduced-order modelling – dimensionality reduction (coordinate transformation), and dynamics discovery – highlighting specifically the role that ML can play in each step.

**4.1. ML for dimensionality reduction and coordinate transformation**

The concept of reduced-order modelling is fundamentally based on the premise that data generated by/for physical systems are usually low-rank. This means that within the often high-dimensional data, there exist a few dominant patterns or structures that are sufficient to explain the system's behavior. These dominant patterns capture the essential dynamics of the system, allowing for a significantly reduced representation without substantial loss of accuracy. Enhancing the coordinate system used to represent the reduced dynamics serves as a major opportunity for integrating ML into ROM development.

To elaborate, we start by recalling from subsection 1.1 that POD and its underpinning concept of SVD provide an orthogonal set of modes, comprising a linear subspace for approximating data [105].

In more detail, SVD decomposes a data matrix $X \in C^{n \times m}$ into three matrices, $U \in C^{n \times m}$, $\Sigma \in R^{m \times m}$, and $V^* \in C^{m \times m}$, hence, $X = U\Sigma V^*$ [106], with the superscript * meaning conjugate transpose. $n$ denotes the dimensionality of the data, and $m$ represents the number of measurements or snapshots. The columns of matrix $U$ are the left singular vectors, which represent the SVD/POD modes. Matrix $V$ contains the right singular vectors. For time-series data, the columns of $V$ represent the temporal pattern. $\Sigma$ is a diagonal matrix with non-negative entries $\sigma_i$, $i = 1,2,...m$ along its diagonal. These $\sigma_i$ entries are the singular values associated with the SVD/POD modes. The singular values are arranged in descending order and represent the "strength" or "importance" of the corresponding singular vectors in the matrices $U$ and $V$ [106].

The above decomposition provides a hierarchical coordinate system with bases defined by the columns of the matrix $U$. Depending on the distribution of singular values and the desired variance to retain within the data, we can truncate the SVD expansion at a certain rank $r$, yielding a low-rank approximation of our data ($\tilde{X}_r$) as $X \approx \tilde{X}_r = \tilde{U}_r \tilde{\Sigma}_r \tilde{V}_r^* = \sum_{i=1}^{r} \sigma_i u_i v_i^*$.

The choice of $r$ is typically guided by examining the distribution of the singular values and/or the normalized cumulative sum of the first $r$ $\sigma_i$ values [106]. A sharp drop in the distribution of singular values indicates that a (relatively) low-rank representation of the data exists, which captures most of the variance. This low-rank representation significantly reduces the dimensionality of the data while preserving its essential features.

Despite the power of SVD to extract low-rank features and patterns from data, many physical systems evolve on nonlinear manifolds. ML and deep learning, thus, present a potent method for generalizing dimensionality reduction from learning linear subspaces as with SVD/POD/PCA to learning coordinates on curved manifolds.

Autoencoder networks [32], in particular, open viable pathways to learn nonlinear data embeddings. Figure 7 shows a schematic representation of an autoencoder with multiple hidden layer (7 in this case). The hidden layers of an autoencoder can include fully connected layers with nonlinear activation functions, convolutional layers, etc.

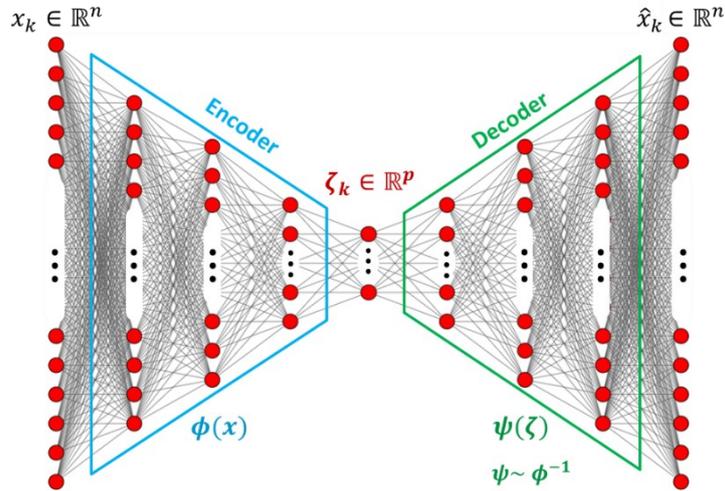

Figure 7: Schematic representation of an autoencoder network. The encoder ($\phi(x)$) reduces the dimensionality of the data ($x_k \in \mathbb{R}^n$), finding the latent space variables ($\zeta_k$) belonging to a reduced subspace $\mathbb{R}^p$, with typically $p \ll n$. The decoder ($\psi(\zeta)$) maps the latent variables of the autoencoder back to the high-dimensional $\mathbb{R}^n$ space. The network is trained by minimizing the loss between the input data to the encoder ($x_k$) and the output data of the decoder ($\hat{x}_k$).

As is shown in Figure 7, autoencoders feature an input and output layer that matches in size the high-dimensional system state. They also include a bottleneck hidden layer that reduces the state to a low-dimensional space, consisting of one or more latent variables ($\zeta$). The encoder part ($\phi$) of the network maps the high-dimensional state ($x_k$, $k$ being a time instant for time-series data, e.g.) to the latent space. Autoencoders belong to the unsupervised learning category. As a result, even though it is the latent space representation at the bottleneck of the autoencoder that we are interested in, the decoder part that reconstructs an estimate of the high-dimensional state ($\hat{x}_k$) from the latent space is crucial for the unsupervised training of the network. A common loss function for the autoencoder architectures is the L2-norm error between the decoder's state reconstruction and the original input state.

A shallow autoencoder with single-layer encoder and decoder as well as linear identity activation functions is closely related to SVD/POD. We will demonstrate this in subsection 4.1.1. In any case, the modular nature of the autoencoders, similar to any neural net, enables expanding the number of layers as well as employing nonlinear activation functions. Hence, the autoencoder can learn nonlinear manifold coordinates, improving compression in the latent space. This is another aspect that the demonstrative example to be presented in subsection 4.1.1 aims to elucidate.

The use of autoencoders have gathered significant attention from across sciences, including in fluid mechanics [107][108]. Lee and Carlberg [109] demonstrated the effectiveness of deep convolutional autoencoders in enhancing classical ROM techniques based on linear subspaces [110]. Additionally, promising data-driven methods for constructing ROMs on spectral submanifolds (SSMs) have been proposed [111].

Clustering and classification are not directly related to dimensionality reduction but are used in innovative ways to aid reduced-order modelling. Classification algorithms, such as the K-nearest neighbors (KNN), can enable detecting and categorizing various dynamical regimes and/or operating conditions of a system. As a result, a potential application may be building a library of ROMs from which a specific relevant ROM compatible with the dynamics regime of the system as represented by the data and distinguished by the classifier would be chosen and used for prediction and/or control.

With regard to clustering, the k-means algorithm has proven effective in various fluid dynamics applications in particular. K-means were employed by Kaiser et al. to develop a data-driven discretization of the high-dimensional phase space for the mixing layer of a fluid flow [112]. Their approach yielded a low-dimensional representation (a small number of clusters). This facilitated the creation of tractable Markov transition models for the flow evolution from one state to another over time. Amsallem et al. [113] used k-means clustering to partition the phase space into distinct regions. Within these regions, local reduced-order bases were constructed, leading to enhanced stability and robustness against parameter variations.

The above utilizations of k-means clustering highlight the potential applicability areas of clustering in general for reduced-order modelling and analysis.

*4.1.1. An example: comparison of the performance of SVD and autoencoders toward dimensionality reduction*

To clarify the previous discussions regarding the SVD and the autoencoders, including their relationship with respect to dimensionality reduction, we compare here the performance of SVD against three autoencoder architectures with an increasing number of fully connected hidden layers and a progressively reduced number of latent variables in the autoencoder's bottleneck layer.

The three autoencoder architectures are schematically illustrated in Figure 8. Through the comparison of the performance, we aim to demonstrate the extent to which the dimensionality of a sample dataset can be reduced using each method and architecture without losing the ability to have an accurate reconstruction of the ground-truth.

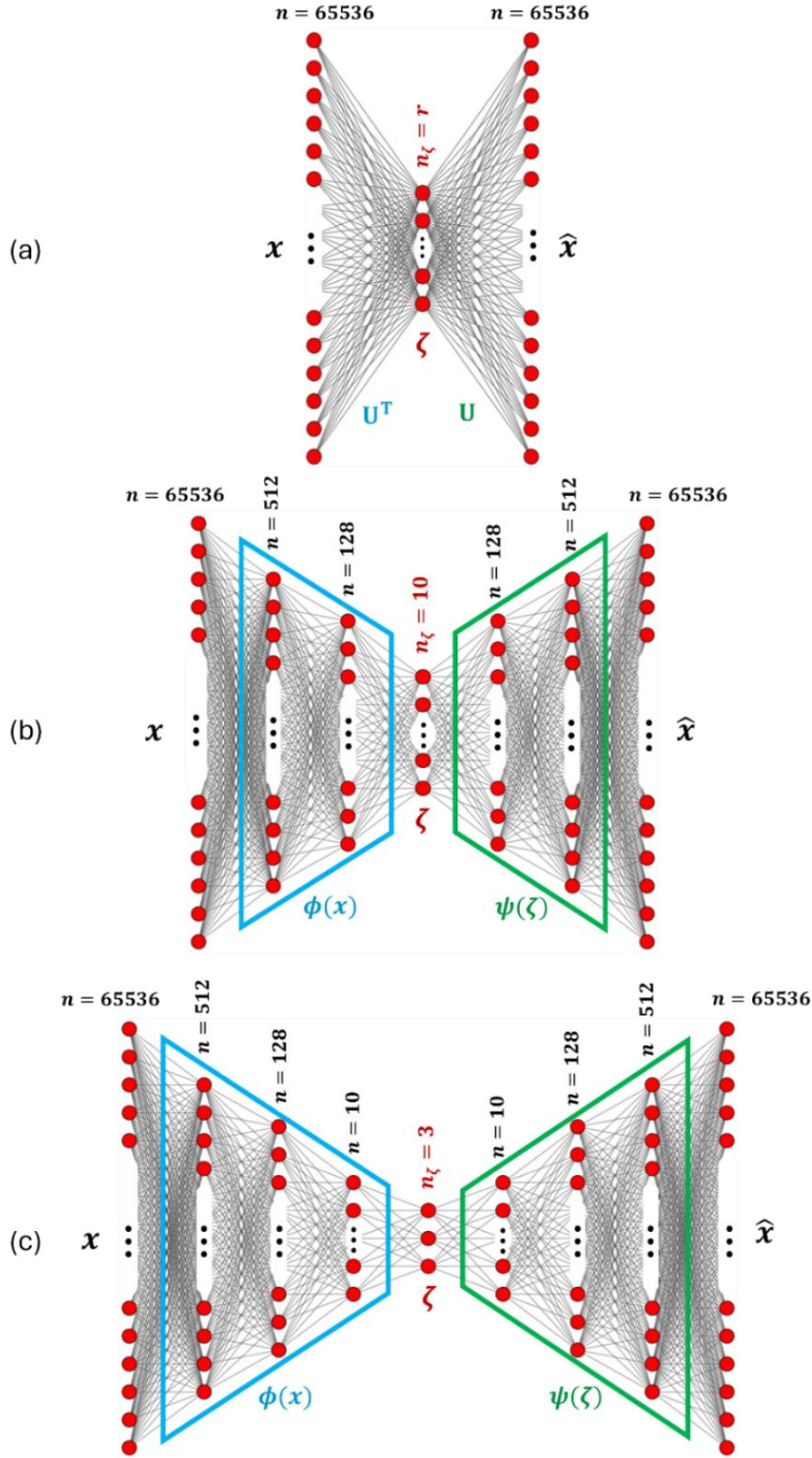

Figure 8: Schematic representation of the autoencoder networks applied to the data from the E×B plasma test case; (a) shallow linear autoencoder, which is equivalent to SVD, (b) autoencoder with 5 hidden layers and 10 latent variables, (c) autoencoder with 7 hidden layers and 3 latent variables.

The example in this subsection is adopted from our previous work [114]. Further details regarding this example, including how the ground-truth data was generated can be found in Ref. [114].

The dataset used for the demonstrations that follow is derived from a 135 $\mu$s-long PIC simulation of a plasma configuration, in which the plasma is subject to mutually perpendicular, externally applied electric (E) and magnetic (B) fields. Such a cross-field or E×B configuration is of great scientific interest and underpins important plasma technologies [2][115], including magnetrons in the plasma manufacturing industry or Hall thrusters for in-space plasma propulsion.

The adopted dataset corresponds to about 8500 snapshots of a specific plasma property – the electrons' axial current density or $J_{ey}$, with $y$ denoting the axial direction in the specific plasma configuration here. The $J_{ey}$ property quantifies the rate of the electrons' cross-magnetic-field motion in the simulated plasma configuration. The total dimensionality of the ground-truth $J_{ey}$ snapshots from the simulation is $n = 256 \times 256 = 65536$. Out the total 8500 snapshots, about 3500 were used for training and about 5000 for testing. Wherever we refer to "test window" in the following discussions, we mean that comparisons are made against any of the 5000 testing (unseen) snapshots.

In the context of dimensionality reduction, testing implies the assessment of how well unseen data can be represented by the identified reduced bases from the dimensionality-reduction algorithm (here, SVD or autoencoder). Accordingly, after determining the reduced coordinates using the training data, the test data are projected onto the reduced coordinates, and then the accuracy of the low-rank reconstructions of the unseen data is evaluated. For SVD, this is mathematically presented as $X_{k,r} = U_r U_r^T X_k$, where subscript $k$ indicates the timestep. For autoencoders, the described projection and reconstruction amount to feed-forward passes of the test data through the trained network.

Figure 9 shows that the reconstructed $J_{ey}$ snapshots using 100 SVD ranks are almost exactly reproducing the ground-truth PIC-simulation snapshots at several random time instants within the test window. A rank-100 representation implies that the dimensionality of the data is reduced by a factor of about 600.

We can also see from Figure 9 that the reconstructed $J_{ey}$ snapshots from the shallow linear autoencoder (Figure 8(a)) with $n_\zeta = 100$ (i.e., 100 latent variables) also provide an excellent reconstruction of the ground-truth and, moreover, represent an identical resemblance to the reconstructed snapshots from SVD. This observation illustrates the point mentioned earlier in subsection 4.1 that a shallow linear autoencoder is equivalent to SVD.

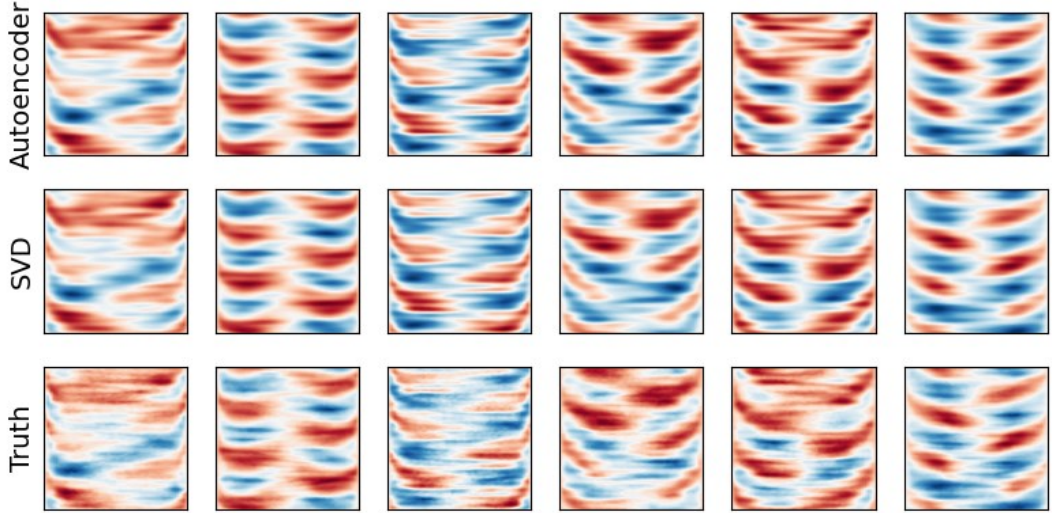

Figure 9: Comparison of the reconstructed normalized 2D snapshots of the electrons' axial current density ($J_{ey}$) from the shallow linear autoencoder (Figure 8(a), **top row**) and the SVD (**middle row**) with $r = n_\zeta = 100$ against the ground-truth snapshots from the PIC simulation (**bottom row**). Each column represents a random sample time instant within the test window.

Since autoencoders can find nonlinear embeddings of the data – a capability enabled by their modularity – architectures with increased hidden-layer numbers and nonlinear activation functions are expected to be able to represent data using a much lower number of latent space variables.

To demonstrate this, we assess in the following the performance of a shallow linear autoencoder with various number of latent variables at the bottleneck. We then show how the reconstruction quality at the extreme of highly low-dimensional latent space improves when we add nonlinearity to the architecture.

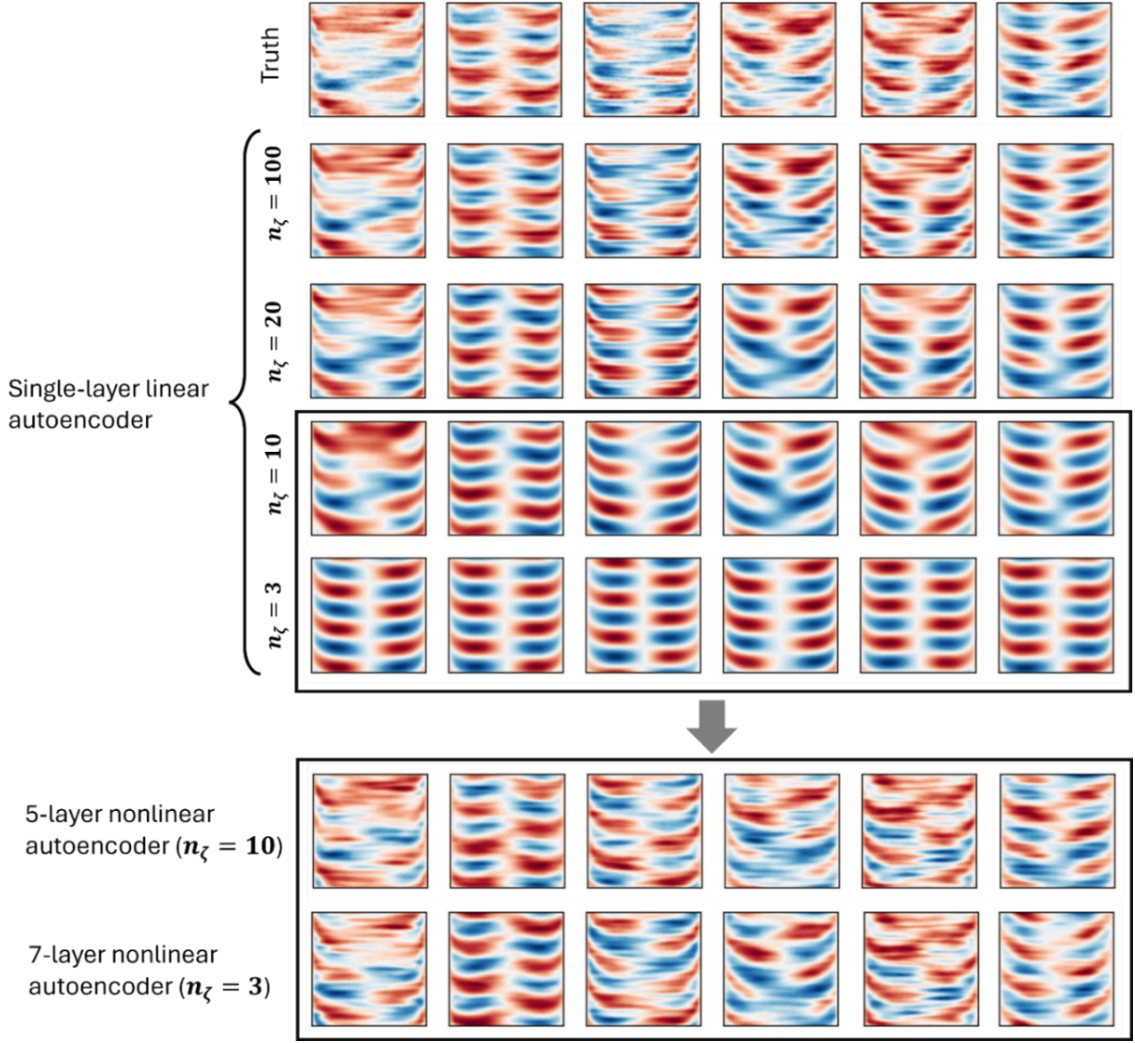

Figure 10: Demonstration of the data reconstruction from compressive representations obtained using the different autoencoder architectures shown in Figure 8 at several random time instants within the test window; (**top subplot**) reconstructions of the $J_{ey}$ data using a single-layer linear autoencoder with varying number of latent variables compared against the ground-truth data snapshots, (**bottom subplot**) $J_{ey}$ reconstructions using nonlinear autoencoders with different hidden layers and latent variables.

Figure 10 shows the snapshots of the $J_{ey}$ data from the PIC simulation and from the autoencoder architectures of Figure 8 at several randomly selected time instants within the test window. Comparing the reconstructed snapshots from the shallow linear autoencoder with various numbers of latent space variables against the ground-truth ones, we notice that, with $n_\zeta = 10$, substantial loss of detailed information in the data occurs, and with merely 3 modes, the reconstruction is inadequate and not at all similar to the ground-truth. These comparisons clearly show that the reconstruction quality of the shallow liner autoencoder (essentially SVD) is poor with a low number of latent variables, failing to capture the essential details of the plasma data.

To address this, we can utilize deeper nonlinear autoencoders. We see from Figure 10 that a 5-layer nonlinear autoencoder can reduce the dimensionality to 10 latent variables while maintaining an accurate representation of the data. The reconstruction quality has significantly improved when using the 5-layer autoencoder with $n_\zeta = 10$ compared to what was observed with 10 modes from the single-layer architecture. The 5-layer autoencoder with $n_\zeta = 10$ has reduced the data dimensionality by more than 6000 times.

By employing a 7-layer nonlinear autoencoder, we can further reduce the dimensionality to 3 while still achieving a remarkably good representation of the data. This degree of compression corresponds to more than 20,000 times reduction in data dimensionality. Indeed, comparing against the ground-truth, the reconstructed snapshots of the $J_{ey}$ from the 7-layer autoencoder are seen to have retained a high level of detail. This demonstrates the capability of sufficiently deep networks to effectively represent complex plasma states in highly compressed latent spaces.

The great efficacy of deep autoencoders in providing a highly reduced nonlinear embedding of the data – as was exemplified here for a sample case of seemingly high-dimensional plasma data – means that these architectures serve as an essential component for developing physics-inspired ML-based ROMs that can be interpretable and readily generalizable.

### 4.2. Data-driven discovery of the dynamics

The discovery of the dynamics, i.e., finding relations and/or governing equations for how the state of a system evolves over time, can also majorly benefit from ML, especially for complex physical systems, such as plasma technologies, for which the complete and closed first-principles system of governing equations may be at least partially unknown.

Data-driven dynamics discovery consists of employing ML/DD algorithms to construct a model for the system's dynamics. This model can take the form of a partial or ordinary differential equation.

There are many ML architectures and data-driven algorithms that can be used for dynamics discovery. Many, if not all, of these architectures and methods can be combined with dimensionality-reduction techniques, such as POD or autoencoders, to discover the dynamics in a suitably identified reduced coordinate system (subspace of underlying modes/structures), hence, yielding computationally efficient ROMs. This in fact corresponds to the framework of simultaneous discovery of the reduced coordinates and the dynamics, which we alluded to within the earlier discussions in Section 4.

A schematic representation of an architecture suitable for this simultaneous learning is illustrated in Figure 11. From this schematic, we note that the coordinate discovery (encoder) part of the architecture reduces the dimensionality of the high-dimensional data at a generic time step $k$ ($x^k$) to a low-dimensional latent space parameterized by the variables $\zeta^k$. The dynamics discovery algorithm now learns in the bottleneck of the autoencoder the time evolution of the latent variables. In a discrete sense, the dynamic discovery algorithm finds the transformation or mapping that advances the variables $\zeta$ one time step to the future ($\zeta^{k+1}$). The decoder part now maps the low-dimensional representation of the predicted future state to the high-dimensional space ($\hat{x}^{k+1}$).

By coupling/pairing the two optimizations corresponding to dimensionality reduction and sparse dynamics identification, we can enforce the network to find a low-dimensional coordinate for the system where the dynamics can be described parsimoniously, i.e., "*as simple as possible but no simpler*" – a compressed version of a quote accredited to Einstein [116]).

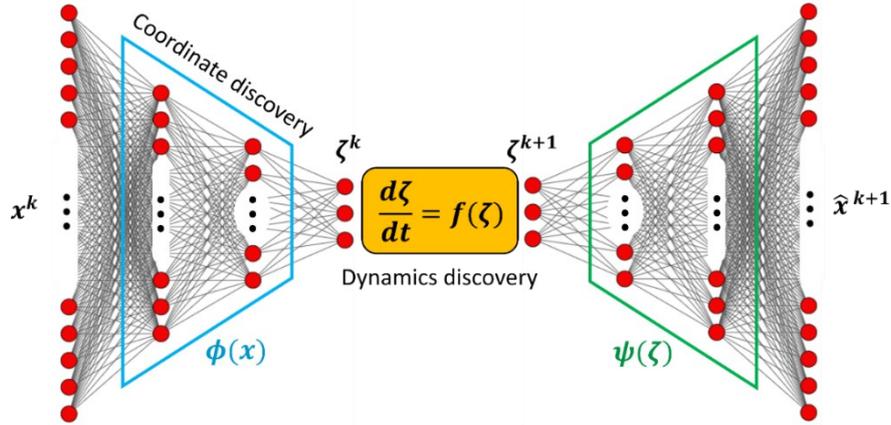

Figure 11: Illustration of a joint dimensionality reduction-dynamics discovery ML architecture for simultaneous coordinate transformation and learning of the dynamics in the reduced coordinates.

With the above remarks in mind, we provide a high-level overview of various categories and examples of linear and nonlinear dynamics discovery algorithms below. In subsection 4.2.1, we will have a closer look at certain promising and interesting methods and architectures. Where applicable, we overview some of the salient prior research efforts that have employed these methods and architectures, alongside presenting demonstrative results from our previous works to illustrate their potential.

Development of linear-dynamics models or linear representations of the nonlinear dynamics are of particular appeal in science and engineering because linear systems offer analytical tractability and there are a wide range of established techniques for their evaluation and analysis. Two well-known data-driven approaches that aim to construct linear time-dynamics models/representations for physical systems are the dynamic mode decomposition (DMD) [117][118] and the more general, closely related framework of Koopman analysis [119].

Various neural network architectures enable effectively capturing nonlinear dynamics. Some notable examples include the LSTM [120] and echo-state networks (reservoir computing) [30] along with GANs [33]. Among these, the applications of GANs have been relatively limited but promising enough to warrant further exploration [121]-[123]. Notably, a GAN-based architecture has been developed and demonstrated recently, which enables reconstructing the full-3D velocity field near the wall of a wall-bounded flow geometry using measurements at the wall [124].

A recent work in plasma physics by Wong et al. [125] has used the reservoir computing (RC) framework to derive ROMs for Hall thruster's dynamics from experimental time-history measurements. The study demonstrated the promise of the RC models to predict the observed behaviors, as well as to infer the values of certain properties from the measurements of the others (see subsection 4.2.1.4 for another example on this type of inference). Wong et al. discussed some caveats to the RC approach as well, including that the derived ROMs are low-fidelity due to the minimal presence of relevant physics and that the influence of the involved network hyperparameters on the predictions are not well understood [125].

Alternative regression techniques exist as well to derive nonlinear dynamical system models. Cluster Reduced-Order Modeling (CROM) [112] has been demonstrated as a straightforward yet robust unsupervised learning method. The CROM approach decomposes time-series data into a few number of representative clusters and then yields models for the transition-state probability between these clusters.

Along a similar direction to CROM and its application within Ref. [112], Kohne et al. [126] employed a data-driven clustering method – the self-organizing maps – to analyze the data from the full kinetic simulations of the plasmoid instability [127], observing that the identified clusters reflect the existing knowledge concerning the characteristics of this instability mode. Even though clustering is not a ROM development method per se, the authors in Ref. [126] pointed out the potential applicability of their method to represent the transition between various models (including ROMs) describing the plasma.

The operator-inference method [128] is another approach to derive nonlinear dynamical system models. It resembles Galerkin projection and aims to learn neighboring operators from data. Furthermore, Sparse Identification of Nonlinear Dynamics (SINDy) [94] learns a parsimonious model by fitting the observed dynamics from data to the fewest terms in a library of candidate functions. Phi Method is a regression-based local-operator discovery algorithm developed recently by the authors [129][130]. Phi Method simultaneously discovers from data the dynamics of the system and the optimal discretization stencil for the identified dynamical terms, providing a discretized system of equations that enables straightforward time forecasting of systems' state [129].

### 4.2.1. Closer review of several dynamics discovery methods

#### 4.2.1.1. *Dynamic mode decomposition*

DMD is a data-driven technique that was introduced by Schmid [117] in fluid dynamics to decompose a flow field into series of coherent spatiotemporal modes with exponential time dynamics. Due to its simplicity of implementation and interpretability, the DMD method has gained remarkable popularity and has been used in various applications, particularly in the field of fluid dynamics [119][131]-[135].

The principal idea behind DMD is to ideally combine a spatial dimensionality-reduction technique, such as POD, with the Fourier transform in time. The method provides a best-fit linear-time-dynamics model describing the spatiotemporal data from a system. Noting these aspects of DMD, we can represent the method as the combination of a single-layer linear autoencoder with a linear operator for the dynamics at the bottleneck (Figure 12). In other words, we can think of DMD as an approach that finds a linear regression model for the data in the reduced SVD/POD coordinates.

Of course, to compute a DMD model for the data, we do not use a shallow autoencoder. However, considering the equivalence between DMD and the architecture shown in Figure 12 provides a tangible example for the discussions earlier in Section 4, as well as in subsection 4.1. The actual approach to derive a DMD model from a system's data is explained in detail in Refs. [106] and [136].

Several variants of the DMD method have emerged over the years, including multi-resolution DMD (mrDMD) [137], high-order DMD (HODMD) [138], and parametric DMD [139].

The recent developments of the optimized DMD (OPT-DMD) [140] and the bagging optimized DMD (BOP-DMD) [141] algorithms have significantly changed the potential of DMD to perform critical modelling tasks due to the robustness of these algorithms to noise. Importantly, OPT-DMD can be used to constrain best-fit linear models to be stable by construction.

Physics-informed DMD (Pi-DMD) [142] aims to constrain the DMD's linear regression of the data by imposing known physics of the problem such as the existing symmetries. The constrained regression underpinning Pi-DMD limit the search space of the optimization to the subspaces (or the manifolds) that respect the symmetry of interest. This has been shown in Ref. [142] to yield significant performance improvements over the basic, unconstrained DMD implementations.

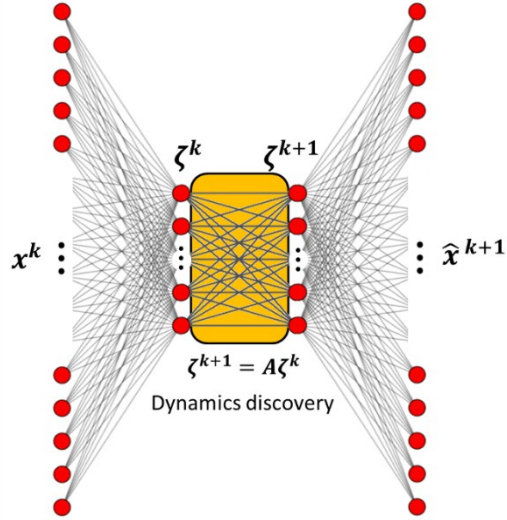

Figure 12: Representation of the DMD method as the combination of a shallow linear autoencoder network and a linear-in-time transformation in the network's latent space.

The DMD method presents major promises for plasma physics applications, in particular with regard to extraction of spatiotemporal coherent structures (instability modes, e.g.) as well as reduced-order modelling. Realizing these promises, several researchers have applied DMD to data from various plasma systems.

Sasaki et al. [143] applied DMD to the simulation data of the Kelvin-Helmholtz instability [144] in order to extract the underlying dominant plasma structures and develop models for the time evolution of these structures (modes). The time dynamics of the DMD modes was calculated using an approach based on "convolution-type correlation integrals" [143]. The authors observed that the derived time dynamics reflects the essential physical behaviors much better that what can be achieved with Fourier analysis [143]. Taylor et al. [145] and Kaptanoglu et al. [146] applied the DMD method to the experimental and simulation data from the HIT-SI fusion experiment. Both studies provided interpretable insights into the dominant patterns underlying the data. Taylor et al. recovered the three dominant modes of the HIT-SI plasma using DMD [145]. The application of DMD by Kaptanoglu et al. [146] enabled uncovering several coherent magnetic modes from the data, which subsequently allowed a previously unrecognized three-dimensional structure in the simulations to be discovered. Maddaloni et al. [147] used DMD to analyze the dynamics of the global discharge oscillations in a Hall thruster plasma using the data from a hybrid fluid/PIC simulation.

DMD is a powerful linear reduced-order modelling approach with great flexibility. A particular example of the DMD's flexibility is that it is readily applicable to data matrices that represent expanded state vectors of physical systems.

In the following demonstrative example, which is extracted from our previous work in Ref. [76], we would illustrate that DMD (the OPT-DMD variant, to be exact) can provide a stable and predictive linear ROM that best describes the time evolution of a multi-property state vector of a plasma system simultaneously.

For this example, OPT-DMD was applied to high-fidelity data from the PIC simulation of the same plasma setup that we described in subsection 4.1.1, i.e., the E×B plasma configuration. The difference here is that the dataset comprised several plasma properties, namely, the electrons' drift velocity components along the radial, azimuthal, and axial directions ($V_{de,x}, V_{de,z}, V_{de,y}$), the radial and azimuthal components of the electron temperature ($T_{ex}, T_{ez}$), and the azimuthal component of the electric field ($E_z$). The OPT-DMD model simultaneously learned the time dynamics of all these properties from training over a window of about 60 $\mu s$ and, in turn, provided a simultaneous prediction of their time variations over a rather extended time window of 45 $\mu s$.

Figure 13 compares the predicted 2D snapshots of the plasma properties from the OPT-DMD ROM against the ground-truth snapshots at three randomly selected time instants within the test window. Across all the time instants and plasma properties, a remarkable degree of similarity between the OPT-DMD-predicted and the ground-truth snapshots is noticed.

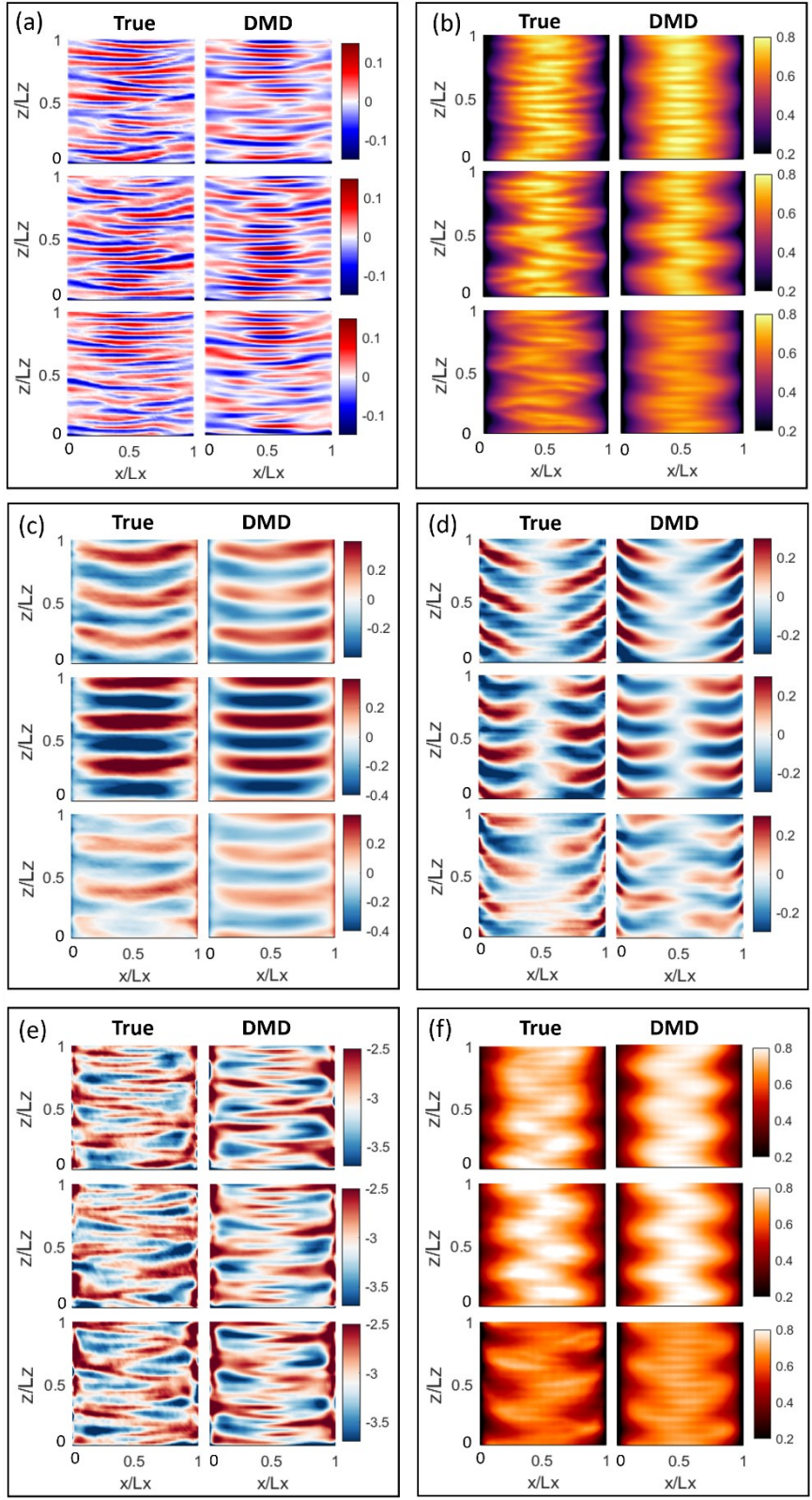

Figure 13: Comparison between the 2D snapshots of several plasma properties from the ground-truth PIC simulation of an E×B plasma discharge (**True** snapshots) and from the OPT-DMD model (**DMD** snapshots) at three randomly selected time instants within the test window with the length of 45 $\mu$s; (a) azimuthal electric field ($E_z$), (b) electron number density ($n_e$), (c) electrons' radial drift velocity ($V_{de,x}$), (d) electrons' axial drift velocity ($V_{de,y}$), (e) electrons' azimuthal drift velocity ($V_{de,z}$), (f) radial electron temperature ($T_{ex}$). Figure adopted from Ref. [76].

It is emphasized that the performance of any ML/DD ROM, including the one from OPT-DMD here, shall be always assessed against the reduction in the model's complexity, as well as the computational cost advantage. Noting this, the accuracy of the predictions from the OPT-DMD ROM are notably good, considering two points: (1) the derived OPT-DMD ROM is a linear-time-dynamics model, describing the time evolution of a plasma system that exhibits highly nonlinear, transient behaviors; (2) the ROM has a negligible computational cost compared to the reference full-order PIC model.

*4.2.1.2. Koopman analysis/transformation*

Many dynamical systems of applied interest, including plasma systems such as the one used for demonstrative purposes in the previous subsection, exhibit nonlinear and/or transient dynamics. The Koopman analysis framework, based on the Koopman theory [148], aims to find a coordinate transformation (embedding) in which the nonlinear dynamics can be linearly represented.

The Koopman operator, a central concept in this framework, involves defining an infinite-dimensional operator that acts on observables (functions) of the state vector from a nonlinear dynamical system. By applying this operator to suitable observables, the dynamics can be described linearly in the observables space. The Koopman operator lifts the dynamics from a finite-dimensional state space to an infinite-dimensional function space, where the evolution of the original system becomes linear.

Despite being an abstraction, the Koopman theory suggests that such a mapping exists, although identifying the appropriate observables to achieve this linear representation is nontrivial and a subject of ongoing research. In this regard, several researchers, e.g. in Refs. [149]-[152], have attempted using deep-learning architectures and autoencoders, like the one illustrated schematically in Figure 14, to identify nonlinear observables and/or coordinate embeddings that may realize Koopman linear representation of nonlinear dynamics.

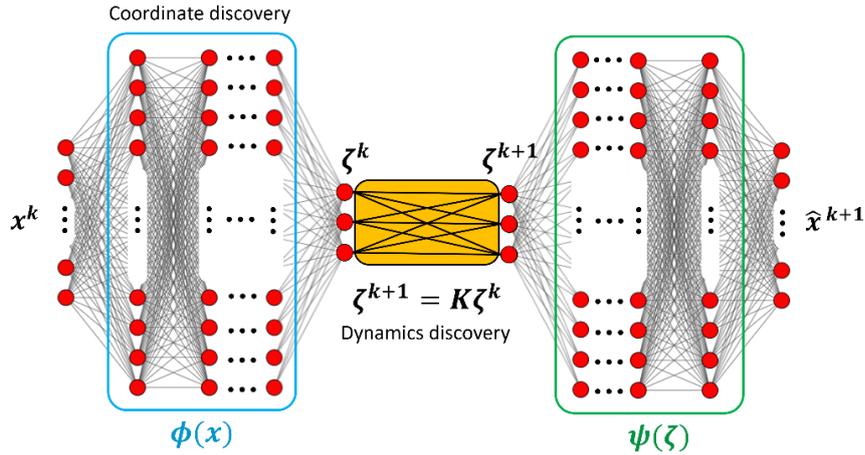

Figure 14: A schematic representation of deep Koopman networks for the simultaneous discovery of the linear embedding of nonlinear dynamics (Koopman transformation) and the associated time-stepping operator $K$.

The DMD method is closely related to the Koopman analysis framework, and efforts have been made to extend DMD with nonlinear observables [153][154] in order to find an approximation for the Koopman operator. The Koopman analysis and the related DMD implementations serve as powerful tools for ROM development of nonlinear systems.

A recent application of Koopman analysis to plasma systems have been made by Carrier et al. in Ref. [155]. In this study, the authors used an autoencoder network with convolutional layers and a Koopman-inspired auxiliary network at the bottleneck to analyze the nonlinear dynamics of a plasma setup. The analyzed setup serves as a representative platform for investigating MHD instabilities common in magneto-inertial fusion devices.

Further to the above, we also provide a demonstrative example here from our prior works [129] to elucidate the connection between the Koopman analysis framework and the DMD method, as well as the utility of these for reduced-order modelling.

To do so, it is first noted that a well-known limitation of the DMD algorithm is its underlying assumption of the linearity of the time dynamics (see chapter 10 of Ref. [118] for more details). However, inspired by Koopman theory, nonlinearities can be introduced into a DMD model by expanding the state vector space to include nonlinear observables (functions) of the state variables, as was done in Refs. [153][154].

The provided example here extends the approach of Refs. [153][154] to a plasma setup. The data were obtained from a PIC simulation of a 1D plasma setup. This setup represents the evolution of a typical Hall thruster discharge along the azimuthal (circumferential) coordinate of the thruster. In this adopted plasma geometry, the azimuthal direction ($z$) is perpendicular to the directions of externally applied, temporally invariant electric and magnetic fields. The electric field is along the axial direction (denoted as $x$ for the specific plasma geometry here), and the magnetic field is along the radial direction ($y$).

The dataset consisted of 4000 snapshots of several output plasma properties (system's state variables), namely, the electron number density ($n_e$), the electrons' azimuthal and axial drift velocity components ($V_{d,ez}$ and $V_{d,ex}$), and the azimuthal electric field ($E_z$). The initial 800 snapshots, which corresponded to the transient evolution of the system before reaching the quasi-steady state (0-2 $\mu s$), were discarded. The adopted OPT-DMD algorithm was then trained over a window that included the snapshots 801 to about 2400 (time duration of ~ 2-6 $\mu s$). The remaining 1600 snapshots, corresponding to the test window of ~ 6-10 $\mu s$, were reserved for verifying the accuracy of the OPT-DMD predictions.

The augmented data matrix, to which OPT-DMD was applied, comprised the rearranged snapshots of the normalized state variables of the plasma system, in addition to some of their nonlinear observables, hence, the following: $n_e$, $V_{d,ez}$, $V_{d,ex}$, $E_z$, $n_e V_{d,ez}$, $n_e V_{d,ez}^2$, and $n_e E_z$.

The time evolution signals of the system's state variables – $n_e$, $V_{de,z}$, $E_z$, and $V_{de,x}$ – at the mid-location of the domain from the OPT-DMD model with nonlinear observables are shown in Figure 15. The time variations are reconstructed during the training interval and are forecasted within the test period. The predicted signals are compared against the ground-truth signals from the PIC simulation. Figure 16 compares the predicted full spatiotemporal evolution of the plasma state variables from the OPT-DMD ROM against the original data.

With reference to Figure 15 and Figure 16, we should first note that the ability of a data-driven ROM to provide reliable predictions of the *local* values of the state variables of a plasma system is inherently a challenging task [129]. This is because the local values of the plasma properties are affected by an array of intricate plasma phenomena and interactions [129]. Furthermore, one must consider that predicting the system's dynamics in the absence of a fully periodic behavior is difficult for a linear-time-dynamics data-driven method like DMD.

Despite these, the OPT-DMD model used here, which incorporates a multi-property state vector with nonlinear observables, is providing reasonably accurate, reliable and stable predictions of the nonlinear dynamics of our plasma system. This emphasizes the great promises of the OPT-DMD approach in conjunction with the Koopman framework within the context of reduced-order plasma modelling and forecast.

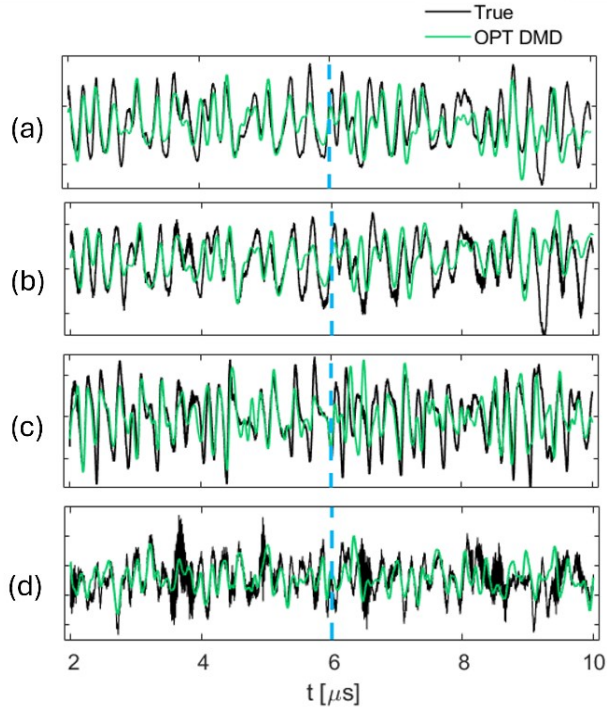

Figure 15: Comparison of the predictions from the OPT-DMD ROM with nonlinear observables against the ground-truth data for the 1D plasma setup; time evolutions of the local values at the mid-location within the simulation domain of **(a)** normalized electron number density ($n_e$), **(b)** normalized electrons' azimuthal drift velocity ($V_{d,ez}$), **(c)** normalized azimuthal electric field ($E_z$), and **(d)** normalized electrons' axial drift velocity ($V_{d,ex}$). Dashed blue lines indicate the end of the training interval. Figure adapted from Ref. [129].

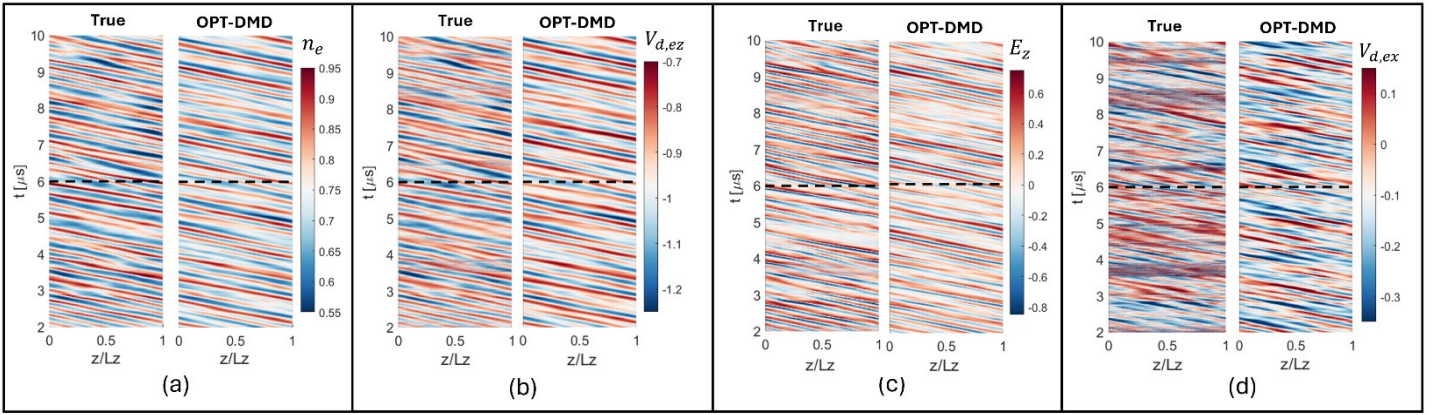

Figure 16: Comparison of the predictions from the OPT-DMD ROM with nonlinear observables against the ground-truth data for the 1D plasma setup; spatiotemporal evolution plots of **(a)** normalized electron number density ($n_e$), **(b)** normalized electrons' azimuthal drift velocity ($V_{d,ez}$), **(c)** normalized azimuthal electric field ($E_z$), and **(d)** normalized electrons' axial drift velocity ($V_{d,ex}$). Dashed black lines indicate the end of the training interval. Figure adapted from Ref. [129].

*4.2.1.3. Evolutionary and library-based regression algorithms*

Machine learning and data-driven techniques are increasingly utilized to develop nonlinear models that strike a balance between predictive accuracy and model complexity. This is meant to mitigate overfitting while ensuring interpretability and generalizability of the derived model.

One such approach for ROM development is "symbolic regression" based on genetic (evolutionary) programming [156]. Genetic programming within symbolic regression works by generating random populations of mathematical expressions, which are then evolved over successive generations through processes such as mutation, recombination, and selection to optimize a specified fitness criterion ("survival of the fittest"). Schmidt and Lipson in 2009 demonstrated genetic programming for the discovery of conservation laws and governing equations [157]. A notable example of the application of symbolic regression within plasma physics modelling is the work of Jorns [83], which was introduced in more detail in Section 3.

Sparse regression on a library of candidate dynamical terms is another well-established dynamics discovery approach, with one of most prominent examples being the SINDy algorithm [94]. SINDy has been employed to identify various dynamical systems [94] as well as partial differential equations [158][159]. Particularly relevant to the paradigm of physics-informed/constrained ML, Loiseau and Brunton in 2018 identified sparse ROMs for various fluid flow systems while enforcing energy conservation as a constraint [160].

Both symbolic regression and sparse [library-based] model identification methods like SINDy utilize Pareto analysis to identify models that offer the optimal trade-off between the predictive accuracy and the model complexity as quantified by the number of included terms. Remarkably, in scenarios where the underlying physics is known, these approaches often uncover the correct governing equations, leading to superior generalizability compared to other prominent machine learning algorithms [16].

We will focus a bit further on SINDy due to the broad interest it has generated. Contrary to the computationally expensive symbolic regression, SINDy offers a more cost-effective alternative for dynamics discovery.

The SINDy algorithm uses time series data to extract dynamical system models best describing the data. As was pointed out above, this is done by performing a sparse regression on a library ($\Theta$) of candidate linear and non-linear terms such that the resulting dynamical system model best approximates the time derivatives ($\dot{X}$) of a system's state variables ($X$) [94]. The sparse regression finds the coefficient values of the minimum number of "active" terms in matrix $\Theta$ [94]. The terms in matrix $\Theta$ can include arbitrary functional forms of the state variables, such as polynomials, spatial derivatives, and/or any combination of these. The derivatives are computed from the data following standard discretization approaches like finite differencing or equivalent.

The derived dynamical systems from SINDy can take the form of PDEs as well as ODEs. In the latter case, SINDy is coupled to an autoencoder (see Figure 17). The autoencoder finds the latent space of the data, and SINDy then learns an ODE in that space which best describes the time evolution of the identified dominant underlying modes [coefficient values ($a_i$s) in Figure 17] [161].

Multiple variants of SINDy have been developed since the algorithm's introduction in 2016 [162]-[167]. Additionally, the approach has been used across the science and engineering applications, e.g., Refs. [168]-[170] in fluid dynamics.

Recently, SINDy has been combined with Bayesian autoencoders in order to augment the simultaneous discovery of the latent space and the dynamics with uncertainty quantification [171].

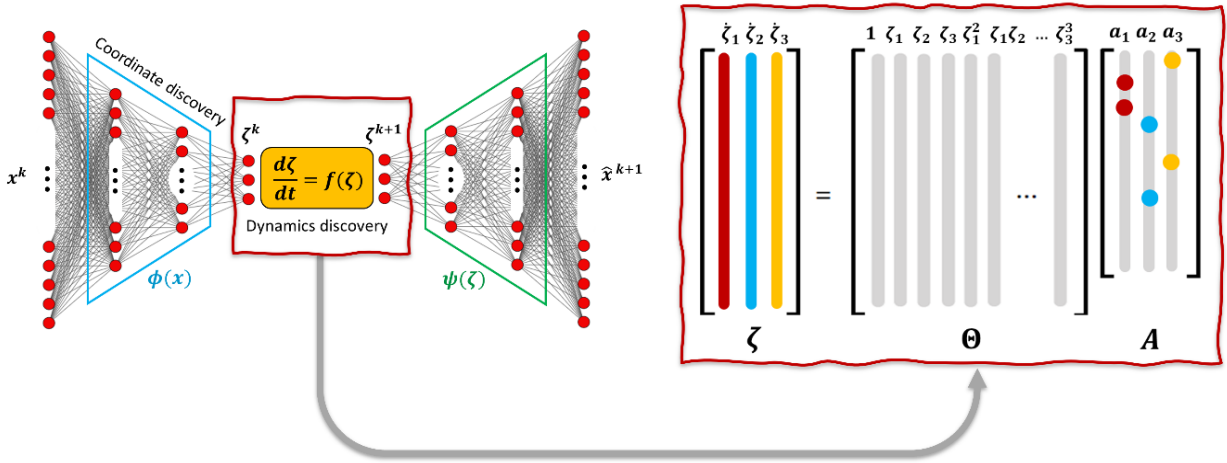

Figure 17: Schematic visualization of the SINDy algorithm as a dynamics discovery method in the latent space of an autoencoder.

SINDy has also gained traction within computational plasma physics research, with a number of recent interesting efforts.

Kaptanoglu et al. investigated first-principles and data-driven approaches toward reduced-order modelling of a class of MHD plasmas, comparing the performance of the classical POD-Galerkin method against POD-SINDy and evaluating the effects of imposing physics constraints on the ROMs [172]. Recently, Kaptanoglu et al. have examined in Ref. [173] the applications of sparse regression methods, such as SINDy, to various tasks in plasma physics, including system identification, compressed sensing, and design optimizations.

The weak (integral) form of the SINDy algorithm was employed by Alvez and Fiuza in Ref. [174] to recover from PIC simulation data the known differential equations, such as the Vlasov and the MHD equations. The authors demonstrated the importance of using their weak-form SINDy implementation so as to robustify the model discovery to the discrete-simulation-particle noise [174]. In Ref. [175], Abramovic et al. applied SINDy to data from the simulations of drift-wave turbulence [176] based on the Hasegawa–Wakatani and modified Hasegawa–Wakatani models. They reported the efficacy of SINDy as a sample sparse regression method to derive the governing nonlinear PDEs accurately describing the evolution of the system [175]. A recent study in Ref. [177] applied the weak-form SINDy for modelling the oscillatory discharge phenomenon of breathing mode [178] in Hall thrusters. The authors found that SINDy could effectively identify interpretable models of various complexity for average plasma properties, consistent with existing zeroth-order models, and qualitatively distinguish subdomains with different dynamics of the plasma [177].

As we mentioned in the high-level overview discussions of subsection 4.2, Phi Method is a recently developed data-driven approach for dynamics discovery that enables learning the discretized form of the differential equations describing a system's dynamics [76][129][130].

Phi Method can be thought of as an extension to SINDy in that it allows simultaneous discovery of the dynamics and the optimum discretization stencil for the involved dynamics variable [129]. The idea of data-driven finding of the optimum discretization stencil in Phi Method is inspired from the general "stencil-learning" research in fluid dynamics (e.g., Refs. [50][51]), which we reviewed in Section 2 with regard to accelerating high-fidelity simulations.

Phi Method has also similarities to library-based regression techniques [153][154][179][180] and the physics-informed DMD [142]. The success and generalizability of Phi Method, however, roots in its constrained regression on a library of candidate terms that is informed by numerical discretization schemes [129].

The demonstrative example we would present in the following before concluding this subsection regards the application of Phi Method for parametric dynamics discovery. Many real-world systems, including plasma technologies, exhibit parametric dependencies in their behaviors/operation. As such, a ML model (approach) that aims to describe (identify) the dynamics of realistic physical systems can only be deemed adequate if it can reliably capture the possible parametric variabilities of the dynamics [130].

The dynamics discovery methods are often limited in terms of adaptability to parametric variations, which is a significant issue on the path of achieving broader generalizability for the resulting dynamics models [130]. "Parameter embedding" is a crucial consideration in data-driven dynamics discovery and model development. It prevents the need to re-train the ML/DD models on new parameter spaces when they are faced with variations in the involved parameters. The library-based nature of Phi Method, which is the sample algorithm in spotlight here, provides a straightforward path for parameter embedding [130].

The parametric dynamical system for the demonstrations that follow is the same plasma system which was introduced and discussed in the previous subsection, i.e., the 1D azimuthal Hall-thruster-representative plasma discharge. The dynamics of this system is strongly dependent on the magnitude of the applied electric and magnetic fields. As a result, training Phi Method on a set of electric and magnetic field values, the aim has been to re-discover from PIC simulation data the parametric PDE that describes the spatiotemporal evolution of a plasma property – electrons' axial flux density ($J_{ex}$) – evaluating the accuracy of the predictions of the parametric Phi Method model over unseen values of the E and B fields.

This problem and the associated results were more extensively discussed in Refs. [130][181]. In terms of the broader perspective, however, some key highlights are discussed here with reference to the plots in Figure 18. The presented plots compare the complete spatiotemporal maps of the $J_{ex}$ as predicted by the parametric Phi Method model against the "true" 2D maps from the PIC simulations for the various test values of the E and the B.

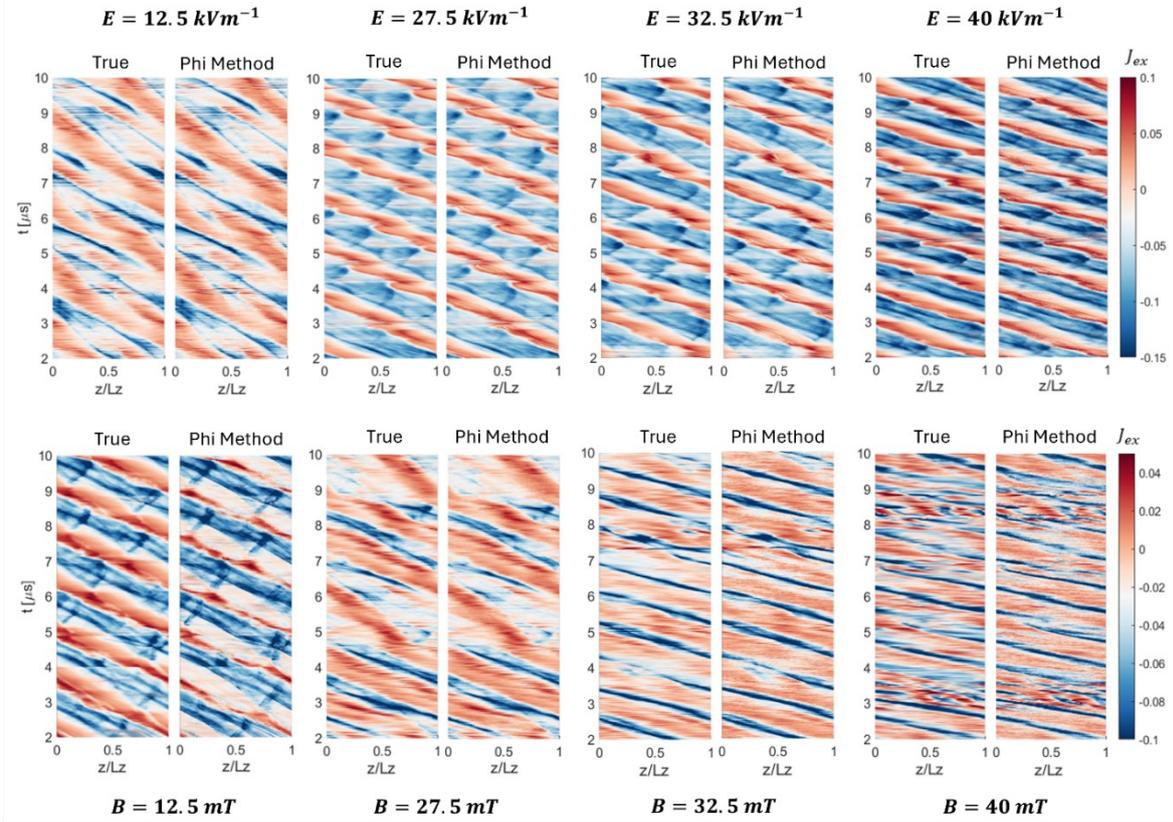

Figure 18: Predictions of the parametric Phi Method model for the 1D azimuthal plasma system compared against the ground-truth data; spatiotemporal maps of the normalized electrons' axial flux density ($J_{ex}$) for the test values of the electric field (E) and magnetic field (B). Figure adapted from Ref. [130]

From Figure 18, the parametric data-driven equation recovered by Phi Method for the $J_{ex}$ is seen to be applicable to across various parameters, which underscores the generalizability of the model. We would point out that the parameter values of E = 12.5 and 40 $kVm^{-1}$ and B = 12.5 and 40 mT fall outside the range of the parameter space over which the parametric Phi Method model had been trained. These cases, thus, represent extrapolation scenarios. However, even though the parametric model's accuracy somewhat degrades, especially against the magnetic field, the predictions are still reasonably aligned with the ground truth.

In Refs. [130][181], the parametric Phi Method model derived from the simulation data was further demonstrated to feature good interpretability in the sense that the equivalent Phi-Method-learned coefficients of the PDE describing the time evolution of the $J_{ex}$ compared rather closely with the analytical coefficients obtained from the finite-difference-discretization of the first-principles PDE.

The provided example serves to emphasize the parametric dynamics discovery as a promising and viable direction to be further explored within the framework of data-driven discovery of the plasma systems' dynamics. In this framework, generalizability and interpretability are crucial aspects to consider.

*4.2.1.4. Shallow Recurrent Decoder (SHRED)*

SHRED is a ML framework that was introduced in 2023 by Williams et al. [77] and Ebers et al. [78]. SHRED optimally combines the power of LSTM networks for dynamics discovery (time-series learning) with the reconstructive advantages of the decoders. The architecture has a broad applicability, including for reduced-order modelling, discrepancy modelling, super-resolution enhancement (see the example in Section 2), and full spatial state inference.

The aim of SHRED is to establish a mapping between the time history of sparse, minimal sensor measurements (at three randomly placed spatial points, e.g.) and the high-dimensional spatiotemporal data [78]. To this end, the LTSM learns the time dynamics from the sensors' trajectory information (time-history measurements), whereas the decoder learns the spatial mapping between the latent output space of the LSTM and the high-dimensional state space of the data.

In practice, the latent outputs of the LSTM are mapped to a compressive representation of the high-dimensional data as encapsulated within the rank-$r$ truncated matrix of right singular values ($V$) from SVD of the data ($X = U\Sigma V^*$) [see subsection 4.1 for more information on SVD]. This mapping to compressive (truncated) representation improves the computational efficiency and the overall performance of the SHRED architecture.

The extracted latent space representation from the LSTM network is agnostic to sensor placement [78], which is a great advantage of SHRED over other widely used methods with similar purposes, such as Shallow Decoder Networks (SDNs) [182]. Moreover, the fully connected nonlinear decoder architecture in SHRED is by construction noise-robust, hence, drastically improving the quality of reconstruction over linear methods, such as POD.

Figure 19 shows a schematic diagram of the SHRED architecture, illustrating how it can reconstruct full spatial maps of various plasma properties from the time history of local probe measurements of a single property – in this case the electron number density ($n_e$). Local density probes collect measurements at three spatial points, which are then processed through an LSTM network. The latent outputs of the LSTM are passed to decoder networks, with each decoder mapping the LSTM's latent variables to the $V$ matrix of the SVD of the data corresponding to a specific plasma property. The reconstructed $V$ matrix for each property is projected to the high-dimensional space using the precomputed SVD modes of that property. As a reminder from subsection 4.1, the SVD modes are represented by the left singular matrix, denoted by $U$.

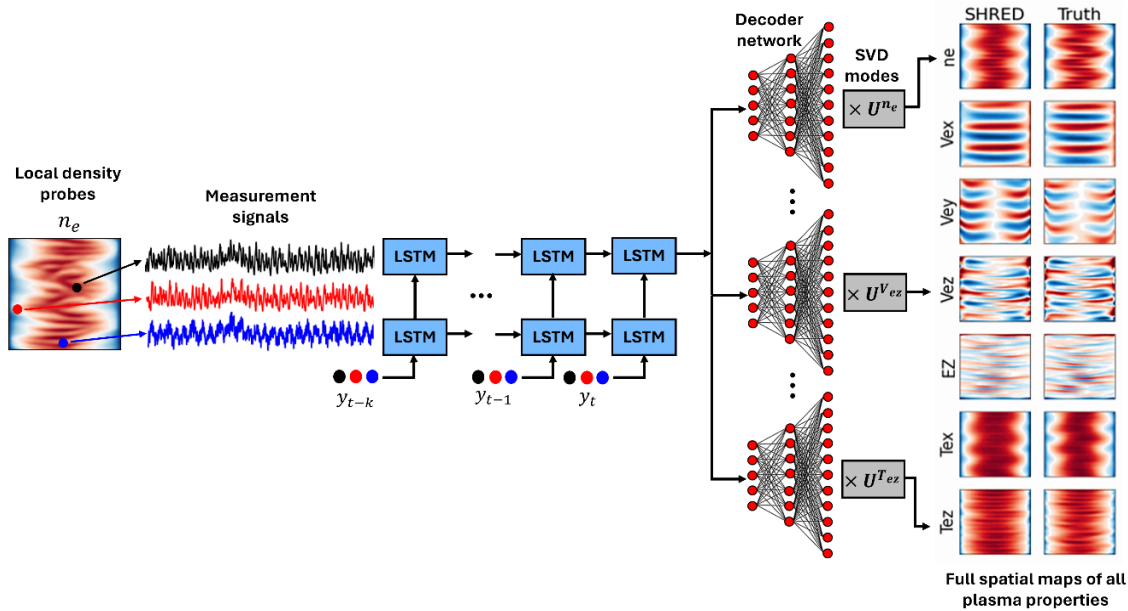

Figure 19: Schematic visualization of the SHRED architecture as applied to data from the PIC simulation of a 2D plasma discharge representative of a radial-azimuthal cross-section of a Hall thruster. Reconstruction of full spatial maps of all plasma properties from the time history of local probe measurements of a single property. The plasma configuration is the same as the one used for the demonstrative examples in Section 2 and subsections 4.1.1 and 4.2.1.1. Figure adopted from Ref. [72].

The SHRED approach can be thought of as the nonlinear ML generalization of the classical method of separation of variables for solving linear PDEs [183]. In this respect, the incorporation of the time-history measurements from the sensors at specific spatial locations into the learning process (system identification problem) serves to make the solution unique, in much the same way that the general form of the solution to a linear PDE from separation of variables is anchored by imposing the initial conditions (local values at *all* spatial locations at *one* time instant).

Noting the above description of the mathematical concept behind SHRED, we would emphasize that, for the kind of mapping visualized in Figure 19, dynamics coupling between the plasma properties (system variables in general) is key. This roots in the fact that a system of $n$ coupled PDEs (each describing the spatiotemporal evolution of one property) can be represented as a single nonlinear PDE of order $n$, to solve which SHRED can be applied.

SHRED has been successfully used for dynamic discovery in a wide range of problems [77][78], including fluid turbulence. It is also recently extended to plasma physics applications [72][183].

In Ref. [77], the authors demonstrated that SHRED could provide predictions of systems' full state based on roll-outs (forecasts) of extremely limited sensor measurements. Demonstrations of similar nature to Ref. [77] but for a representative cross-field plasma system are provided in Ref. [183].

SHRED is highly flexible with respect to the nature of the time-history measurements it requires to learn the dynamics of the system and produce ROMs, inferring the full spatiotemporal state of the system in the process. For instance, the demonstrative example below through Figure 20 shows that SHRED can work equally well with input signals that represent the time evolutions of global (averaged) plasma properties rather than being local measurements of a plasma field.

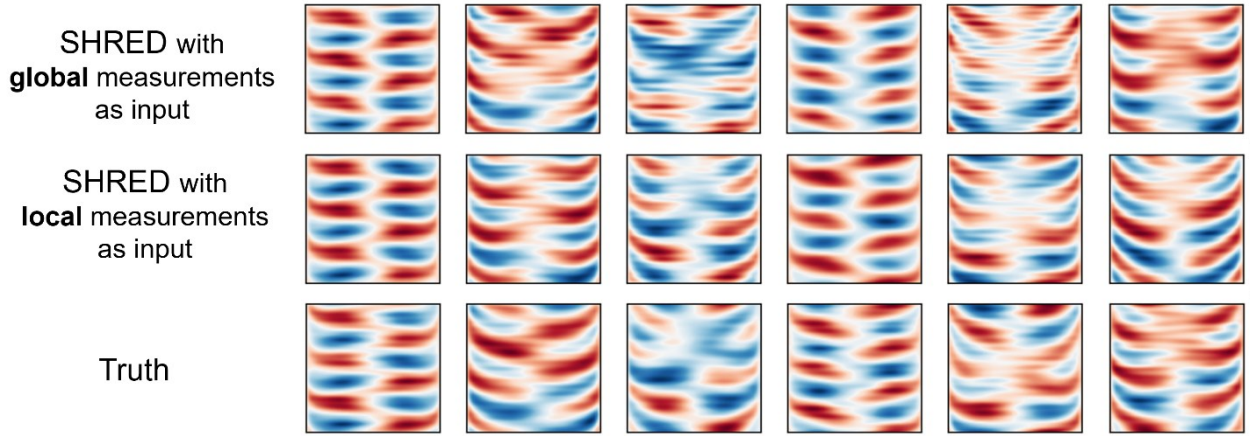

Figure 20: Comparison of the inferred normalized 2D snapshots of the electrons' axial current density from SHRED with measurements of global (spatially averaged) plasma properties as inputs (**top row**) and from SHRED with local measurements as inputs (**middle row**) against the corresponding ground-truth snapshots from the PIC simulation of the 2D cross-field plasma configuration (**bottom row**). Each column corresponds to a randomly selected time instant within the test window.

The data for the example here correspond to the same 2D plasma configuration and the associated PIC simulation, which we referred to in Section 2 and subsections 4.1.1 and 4.2.1.1 as well. The inputs to SHRED were the spatially averaged (mean) values of three selected plasma properties – the electron number density ($n_e$), the electrons' radial current density ($J_{ex}$), and the electrons' axial current density ($J_{ey}$). With these three time histories of global properties, we reconstructed the full spatial distribution of the $J_{ey}$ over time. As another comparative point of reference besides the ground-truth data, Figure 20 also presents the results from full spatiotemporal dynamics discovery of the $J_{ey}$ using three local (virtual) probe measurements of the same property as inputs to SHRED.

Figure 20 illustrates that, across all the six random test-window time instants for which the snapshots from SHRED and the PIC simulation are shown, the spatiotemporal dynamics learned by SHRED using global measurements as inputs are remarkably similar to the ground-truth, as good as the reconstructions using three local sensor measurements.

The presented discussions and sample results highlight the flexibility of SHRED as suggested earlier. They additionally emphasize its promise for dynamics discovery and state-space inference, particularly in settings of applied relevance, where some information from the system are often available and could be combined with the predictions of plasma simulations to create reliable and robust ML ROMs that accurately represent the real-world processes.

**Section 5: Concluding remarks**

We would underline a few important remarks in this final section of the present Perspective.

First, despite the great promises that ML offers for science and engineering, there are still several challenges to be addressed so as to realize the full potential. Some of the most important challenges regard the amounts of data needed, the robustness of the ML algorithms to work with often noisy and sparse data, and the interpretability and generalizability of the ML models.

The data from complex physical systems of real-world applied interest are commonly quite expensive to obtain either numerically or experimentally. As such, the datasets are typically rich only along a single dimension. For example, high-fidelity simulations' results typically feature high resolution in space but are sparse in time. Contrarily, experimental data can cover long time durations but with low spatial resolution and often high levels of noise. It is, thus, essential to develop and mature ML algorithms that can robustly and reliably work with limited and noisy data.

Interpretability and generalizability are also crucial for most scientific and engineering applications, especially in scenarios where decision making is involved based on the model's outcomes/predictions. Interpretability facilitates understanding the workings of a model, and generalizability can significantly lower the cost of its development by expanding the range of conditions and situations to which the single ML model is applicable. Toward addressing many of these challenges, the incorporation of the known physics of a system in terms of, e.g., the conservations, symmetries, and invariances, has provided an important and viable pathway [32]. This is why the theme of "physics-informed" [37] and, more broadly, "physics-consistent" [16] machine learning is being rigorously researched today, and why we advocated for the framework of "physics-inspired" ML in Section 4.

Second, it is true that the interest in ML applications for plasma modelling and simulation has surged over the past years. This rise in community attention has been accompanied by the emergence of a range of research efforts. As was seen through this paper, these efforts can be mainly categorized into enhancing/accelerating plasma simulations, on one side, and system identification (dynamics discovery) and reduced-order modelling, on the other side. Some researchers have also attempted employing ML methods and NN architectures to enhance scaling laws for plasma technologies, such as Hall thrusters [184]-[186], or to facilitate their performance optimization [187].

Nevertheless, ML applications in plasma physics can be still considered as nascent and relatively limited in scope. This is particularly in view of the trends in fluid mechanics – the closely related, adjacent field to plasma physics – which has seen a major, exponential increase in ML-enabled applications for a wide variety of purposes in recent years. Despite the existing gap, the significant progress in ML-related applications within fluid mechanics together with the notable overlaps in tools and practices that exist between fluid mechanics and plasma physics provide excellent opportunities for the cross-transfer of knowledge and capabilities in terms of ML algorithms and use cases to plasma physics. In fact, the mapping of the promising ML applications already demonstrated for fluid flows modelling to plasmas was a main objective of this Perspective.

At this point, one may wonder what reason(s) might have led to the creation of the present gap between the two adjacent fields of fluid mechanics and plasma physics in terms of the prevalence of ML applications. This brings us to the third important remark here. Different explanations may be provided for the created disparity. However, a fundamental root cause can be traced to the scarcity of large enough volumes of data from plasma systems to support ML research for plasma physics.

In this regard, the complexities associated with the plasmas behavior, including the greatly multiscale, multidimensional spatiotemporal characteristics of the involved dominant phenomena, has made the generation of sufficiently varied, statistically representative, and high-quality data a grand challenge in plasma physics. On the one hand, the high-fidelity numerical tools have been largely computationally costly to simulate the real-world geometries of plasma devices over their large relevant parameter spaces of operation. On the other hand, plasma experiments have been highly complicated due to various challenges including accessibility and the cost of operation of the devices, in addition to issues concerning the spatiotemporal resolution and sensitivity of the diagnostics.

As a result, whereas "big data" has become a reality in fluid mechanics over the past decade due to extensive efforts and progress in high-performance computing and experimental measurement capabilities [16][188], similar advances have just begun to manifest and influence the plasma physics research, meaning that plasma physics is still away from the realm of "big data" so that ML can be exploited in the field to its fullest potential.

In any case, community-wide efforts and coordination initiated within plasma physics communities in the past years to address the computational cost issue of high-fidelity plasma simulations, as well as to improve the employed diagnostics, have been changing the circumstances. For example, in the context of computational plasma physics tools, several efforts in recent years have been aimed at tackling the enormous computational burden of the PIC simulations. These efforts include the development of implicit and/or energy-conserving PIC algorithms [65][189], GPU implementation of the PIC method [190], the sparse-grid PIC scheme [191][192], and the reduced-order PIC scheme [73][193]-[196].

Accordingly, ML research for plasma physics is today greatly timely and indeed essential for further advancements of the field.

Lastly, one of the most relevant – and perhaps strong motivators for – industrial applications of physics-informed ML and the ML/DD ROMs is the transformative technology of digital twins (DTs) [4][80]. This digital technology has seen a surge in its appeal across the industries over the past few years along with the significant increase in computational power and the rapid advances in ML algorithms and diagnostics/sensing capabilities that all serve to strongly empower the DTs.

The DT concept extends beyond the conventional notion of simulation and modelling, which are considered static, by the merit of a bidirectional interaction between the digital and the physical twin [4][80]. This bidirectional interaction allows the dynamic evolution of the DT and comprises feedback flows of information from the physical system to the virtual representation and from the virtual back to the physical system to enable decision-making and control. A DT is particularly powerful for the "what-if scenario" analysis in a way that far exceeds what is feasible/practical to do through experiments alone [4][80].

The promises of the DT technology for the plasma industry revolve around two pillars [4][80]: **(i)** streamlined, efficient, and methodical computer-aided design and development processes, **(ii)** robust autonomous operation and control of the plasma system.

The DT technology is closely aligned with and in fact build upon predictive plasma models. However, because of the continual exchange of data that exists between a physical system and its digital twin, the DT can transcend and expand the capabilities and use cases of any predictive plasma model. Moreover, as the DT aims to replicate the behaviour of its physical twin as closely as possible (ideally, in an identical manner) and because the integration of physics-based/informed predictive models under the hood of a DT enables, in principle, extensive interrogation and probing of the processes ongoing during the device's operation, DTs also open novel and unprecedented pathways for scientific inquiry.

Accordingly, in the authors' opinion, the pursuit of fully fledged DTs for plasma technologies provides an optimal overarching framework of both academic and industrial relevance for the ML research in plasma physics. ML/DD ROMs that can cost-effectively and accurately describe the plasma systems' time dynamics by learning from the simulation and/or experimental data are in fact essential enabling elements of the DT technologies.

In addition, the concept of discrepancy modelling (see Section 3) and the role of ML in increasing the representativeness between the digital representation models and the physical asset based on a stream of real-world data serve as crucially important and pertinent lines of research.

**Acknowledgments**:

The authors would like to express their gratitude to Professor Nathan Kutz for the fruitful discussions that helped framing and refining this Perspective. We also thank Dr. Aaron Knoll for his support during the preparation of this work, as well as his contributions to previous works cited within this paper.

FF and MR acknowledge the PhD scholarships received from Imperial College London's Department of Aeronautics, which supported part of this research.

**Data Availability Statement**:

No new data has been produced and/or analyzed for this work.